\documentclass[12pt]{article}
\usepackage{amsfonts,amssymb,amscd,amstext}
\usepackage{a4wide,theorem}
\usepackage{float}
\usepackage{graphicx}
\usepackage{verbatim}
\makeatletter
\@addtoreset{equation}{section}
\@addtoreset{figure}{section}
\@addtoreset{table}{section}
\makeatother

\floatstyle{ruled}
\restylefloat{table}

\newcommand\ie{\emph{i.e.}}

\newcommand\eg{\emph{e.g.}}
\newcommand\g{\ensuremath{\mathcal{G}}}

\newcommand\be{\begin{equation}}
\newcommand\ee{\end{equation}}

\def\fijk{{f_{ij}}^k}
\def\rtth{\sqrt{3}}

\def\fquad{\, , \quad \quad \quad \quad}

\newcommand\Omis{{\Omega^{(2m-1)}_{i_1  \ldots i_{2m-2}}}^\sigma}

\newcommand\tmjfi{\frac{1}{(2m-2)!}}
\newcommand\smj{s_{2m-2}}
\newcommand\smjb{\overline{s}_{2m-2}}
\newcommand\Wmj{W_{2m-2}}
\newcommand\Pzw{P^{(0)}_W}
\renewcommand\S{\cal{S}}
\newcommand\K{\mathcal{K}}
\newcommand\Km{\mathcal{K}_{2m-2}}

\newcommand\C{\mathcal{C}}

{
\theoremstyle{break}
\theorembodyfont{\normalfont}
\newtheorem{theorem}{Theorem}[section]
\newtheorem{corollary}{Corollary}[section]
\newtheorem{lemma}{Lemma}[section]

\newtheorem{definition}{Definition}[section]
}

\newenvironment{remark}{\list{}{\setlength{\leftmargin}{0pt}}\item\relax
\noindent\emph{Remark.~}}{\endlist}

\newenvironment{proof}{\list{}{\setlength{\leftmargin}{0pt}}\item\relax
\noindent\emph{Proof.~}}{\endlist}

\begin{document}

\renewcommand{\thefootnote}{\fnsymbol{footnote}}
\thispagestyle{empty}

\vspace*{-1cm}

\begin{flushright}
FTUV/98-62, IFIC/98-63, DAMTP 1998-141
\end{flushright}
\vskip 2cm

\begin{center}
\begin{Large}
\bfseries{Higher order BRST and anti--BRST operators and cohomology
for compact Lie algebras}
\end{Large}

\vspace*{0.6cm}

\begin{large}
C.~Chryssomalakos$^1$, J.~A.~de~Azc\'arraga$^1$, A.~J.~Macfarlane$^2$
and
\\
J.~C.~P\'erez Bueno$^1$\footnote{
e-mail addresses: chryss@lie.ific.uv.es,
azcarrag@lie1.ific.uv.es, a.j.macfarlane@damtp.cam.ac.uk,
\\
pbueno@lie.ific.uv.es}
\end{large}

\vspace*{0.6cm}
\begin{it}
$1$ Departamento de F\'{\i}sica Te\'orica, Universidad de Valencia
\\
and IFIC, Centro Mixto Universidad de Valencia--CSIC
\\
E--46100 Burjassot (Valencia), Spain
\\[0.4cm]
$2$ Department of Applied Mathematics and Theoretical Physics,\\
Silver St., Cambridge, CB3 9EW, UK
\end{it}

\end{center}
\vspace*{2cm}
\begin{abstract}
After defining cohomologically higher order BRST and anti-BRST 
operators for a compact simple algebra \g, 
the associated higher order Laplacians are introduced and
the corresponding supersymmetry algebra $\Sigma$ is analysed.
These operators act on the states generated by a set of fermionic ghost fields 
transforming under the adjoint representation.
In contrast with the standard case, for which the Laplacian is given
by the quadratic Casimir, the higher order Laplacians $W$ are not
in general given completely in terms of
the Casimir-Racah operators, and may involve the ghost number operator.
The higher order version of the Hodge
decomposition is exhibited. 
The example of $su(3)$ is worked out in detail, including the
expression of its higher order Laplacian $W$.
\end{abstract}

\setcounter{footnote}{0}
\renewcommand{\thefootnote}{\arabic{footnote}}

\newpage


\section{Introduction}
\label{Intro}
BRST symmetry~\cite{Bec.Rou.Sto:76,Tyu:75}, or `quantum
gauge invariance', has played an important role
in quantisation of non-abelian gauge theories.
The nilpotency of the operator $\mathcal{Q}$ generating the global BRST symmetry
implies that the renormalisation of gauge
theories involves cohomological aspects:
the physical content of the theory belongs to the kernel of $\mathcal{Q}$, 
the physical 
(BRST invariant) states being defined by BRST--cocycles modulo BRST--trivial 
ones (coboundaries).
The inclusion of the BRST symmetry in the Batalin--Vilkovisky 
antibracket--antifield formalism (see \cite{Hen.Tei:92,Gom.Par.Sam:95} for 
further references), itself of a rich geometrical structure 
\cite{Hen.Tei:92,Gom.Par.Sam:95,Wit:90,Khu.Ner:93,Ale.Kon.Sch.Zab:97,Ner:95,
Bra.Hen.Wil:98}, and
where the antifields are the sources of the BRST transformations, has made of 
BRST quantisation the most powerful method for quantising systems possessing 
gauge symmetries.
In particular, it is indispensable for understanding the general structure of 
string amplitudes.
It is thus interesting to explore its possible generalisations  and their 
cohomological structure.

An essential ingredient of $\mathcal{Q}$ is what we shall denote here
the BRST--\emph{operator}
\begin{equation}
s = -{1\over 2} {C_{ij}}^k c^i c^j {\partial \over \partial c^k}
\quad,\quad
i=1,\ldots ,r=\dim \g
\quad ,
\label{exprs}
\end{equation}
where the $c^i$ are anticommuting Grassmann
(or \emph{ghost}) variables transforming under the adjoint representation
of the (compact semisimple) Lie group $G$ of Lie algebra \g.
In Yang Mills theories the $c$'s correspond
to the ghost fields, and~(\ref{exprs}) above is just part of 
the generator of the BRST transformations for the gauge group $G$.
This paper is devoted to the generalisations of
(\ref{exprs}) and its associated \emph{anti}--BRST \emph{operator}
$\overline{s}$~\cite{Alv.Bau:83,Mul:86,Ger:86,Bar.Yan:87,Hol:90,
Bau:95,Yan.Lee:96}.

Using Euclidean metric to raise and lower indices\footnote{
The Killing tensor $k_{ij}$ is proportional to $\delta_{ij}$ 
since \g\ is compact and semisimple.}
$\overline{s}$ is given by
\begin{equation}
\overline{s}=\frac{1}{2} {C_{ij}}^k c_k
{\partial \over \partial c_i}{\partial \over \partial c_j} \, \, . 
\label{exprsb}
\end{equation}
The $s$ ($\overline{s}$)
operator increases (decreases)  the ghost number by one. The BRST and
anti--BRST operators may be used to construct a 
Laplacian~\cite{Kaw:86,Ger:86,Bar.Yan:87,Fig.Kim:88,Hol:90,Yan.Lee:96}, 
$\Delta= \overline{s}s+s\overline{s}$;
clearly, $\Delta$ does not change the ghost number.
It turns out (see~\cite{Ger:86,Bar.Yan:87,Hol:90}) that this 
operator
is given by the (second order) Casimir operator of \g.

A few years ago, van Holten~\cite{Hol:90} discussed the BRST complex,
generated by the $s_\rho$ and $\overline{s}_\rho$ operators,
\begin{equation}
s_\rho= c^i \rho(X_i) - {1\over 2} {C_{ij}}^k c^i c^j {\partial \over
\partial c^k} \,  
\fquad \quad 
\overline{s}_\rho = -\rho (X_i){\partial \over \partial c_i} 
                     + \frac{1}{2} {C_{ij}}^k c_k
{\partial \over \partial c_i}{\partial \over \partial c_j} \, \, ,
\label{ssbexpr}
\end{equation}
in connection with the cohomology of compact semisimple Lie algebras.
They act on generic states of ghost number $q$ of the form
\begin{equation}
\psi= {1\over q!} \psi^A _{i_1\dots i_q} c^{i_1}\dots c^{i_q}
\otimes e_A \quad .
\label{psigen}
\end{equation}
The operators in~(\ref{ssbexpr}) 
differ from those in~(\ref{exprs}), (\ref{exprsb}) by the inclusion
of the $\rho$ term, where ${\rho^A}_B$ is a representation of the
Lie algebra \g\ on a vector space $V$ with basis $\{e_A\},  \, A=1,
\dots , \dim V$. However (see
the \emph{Remark} in section~\ref{SBchs}), only the trivial 
representation case is interesting. For $\rho$=0 the 
generic states have the form

\begin{equation}
\psi = {1 \over q!} \psi_{i_1 \ldots i_q} c^{i_1} \ldots c^{i_q} \quad ,
\label{BRSTCOC}
\end{equation}
and we shall consider mostly this case.
The operators $s$, $\overline{s}$ and the Laplacian $\Delta$
may be used to define $s$--closed, $\overline{s}$--closed (coclosed)
and harmonic states. A state $\psi$ (eqn. (\ref{BRSTCOC})) 
is called $s$--closed, $\overline{s}$--closed or harmonic if $s\psi=0$,
$\overline{s}\psi=0$ or $\Delta\psi=0$ respectively. In this way, 
and using the
nilpotency of $s$ and $\overline{s}$, one may introduce
a Hodge decomposition for (\ref{BRSTCOC}) as a sum of an $s$--closed,
a $\overline{s}$--closed and a harmonic state. The interesting 
fact is that, using the above Euclidean metric on \g, 
one may introduce a positive scalar product among states $\psi'$, 
$\psi$, of ghost numbers $q'$, $q$ by
\begin{equation}
\langle \psi',\psi \rangle := {1\over q!}
\psi'_{j_1\dots j_q} \psi^{j_1 \dots j_q} \delta_{q'q}
\quad .
\label{scalprod}
\ee 
Using the Hodge $*$ operator for $\delta_{ij}$ it 
follows that $s= (-1)^{r(q+1)}*\overline{s}*$ (on states
of ghost number $q$), and that  $s$ and  $\overline{s}$
are also adjoint to each other with respect 
to the scalar product~(\ref{scalprod}). As a result, there is
a complete analogy between the harmonic analysis of forms, in
which $d$ and $\delta= (-1)^{r(q+1)+1} * d *$ are adjoint to each 
other, and the
Hodge-like decomposition of states $\psi$ for the
operators $s$, $\overline{s}$~\cite{Hol:90} (see also 
section~\ref{SBchs}). This follows from the fact 
that, due to their anticommuting character,
the ghost variables $c$ may be identified~\cite{Bon.Cot:83}
with (say) the left-invariant one--forms 
on the group manifold $G$, so
that the action of $s$ on $c$ determines the Maurer-Cartan
(MC) equations.

The nilpotency of the BRST operators (\ref{exprs}) or (\ref{ssbexpr})
results from the Jacobi identity satisfied by 
the structure constants ${C_{ij}}^k$  of \g.
This identity can also be viewed as a three-cocycle condition on 
the fully antisymmetric ${C_{ijk}}$, which define 
a non-trivial three-cocycle for any semi-simple \g.
This observation indicates the existence of a generalization
by using the higher order cocycles for \g.
The cohomology ring of all compact simple Lie algebras of rank $l$ 
(for simplicity we shall assume $G$ simple henceforth)
is generated by $l$ (classes of) non-trivial primitive cocycles,
associated with the $l$ invariant, symmetric 
primitive polynomials of order $m_s$, ($s=1, \dots ,l$)
which, in turn, define the $l$ Casimir-Racah 
operators (\cite{Rac:50,Gel:50,Per.Pop:68}; see 
also \cite{Azc.Mac.Mou.Bue:97} and references therein).
The different integers $m_s$ depend (for $s\neq$1)
on the specific simple algebra considered. 
It has been shown in~\cite{Azc.Bue:97} that, 
associated to each cocycle of order $2m_s-1$ 
there exists a higher order BRST operator $s_{2m_s-2}$ carrying ghost 
number $2m_s-3$, defined by the coordinates $\Omega_{i_1\dots i_{2m_s-2}}^j$
of the ($2m_s-1$)-cocycle on \g\
(we shall also give in (\ref{sconro}) the 
corresponding operator $s_{\rho\,(2m_s-2)}$ for the $\rho\neq 0$ case).
The $\Omega_{i_1\dots i_{2m_s-2}}^j$
may also be understood as being the (fully antisymmetric) higher 
order structure constants of a higher ($2m_s-2$) order 
algebra~\cite{Azc.Bue:97}, for which
the multibrackets have ($2m_s-2$) entries.
The standard (lowest, $s$=1) case corresponds to $m_1=2$ $\forall$
\g\ (the Cartan-Killing metric), to the three-cocycle $C_{ijk}$ 
and to the ordinary Lie algebra bracket.
The ($2m_s-2$)-brackets of these higher order algebras satisfy 
a generalized Jacobi identity which again follows from the fact that the
higher order structure constants define ($2m_s-1$)-cocycles for the Lie
algebra cohomology. These ($2m_s-2$)-algebras 
constitute a particular example (in which only one coderivation
survives) of the strongly homotopy algebras \cite{Lad.Sta:93},
which have recently appeared in different physical theories 
which share common cohomological aspects, as in closed string field 
theory~\cite{Wit.Zwi:92,Lia.Zuc:93}, the higher order generalizations
of the antibracket~\cite{Ber.Dam.Alf:96,Akm:97} and the Batalin--Vilkovisky 
complex (see \cite{Sta:97} and references therein).
Higher order structure constants satisfying generalized Jacobi identities of 
the types considered in \cite{Azc.Bue:97} (see (\ref{SGJI}) below) and 
\cite{Azc.Izq.Per:98} have also appeared in a natural way in the extended 
master equation in the presence of higher order conservation laws 
\cite{Bra.Hen.Wil:98}.

In section~\ref{HoBo} of this paper we introduce, 
together with the $l$ general BRST 
operators for a simple Lie algebra, the corresponding $l$ 
anti--BRST 
operators $\overline{s}_{2m_s-2}$ and their associated higher order 
Laplacians. We show there that harmonic analysis 
may be carried out in general (the standard case in \cite{Hol:90}
corresponds to $m_s=m_1=2$), although the 
Laplacians do not in general  correspond to the Casimir-Racah
operators. Nevertheless, we shall show that
${s}_{2m_s-2}$ and $\overline{s}_{2m_s-2}$ are related to
each other by means of the Hodge $*$ operator, and that
they are also adjoint of each other. After showing that
the different higher order BRST, anti-BRST and Laplacian operators
generate, for each value $s=1,\dots,l$, a supersymmetry algebra $\Sigma_{m_s}$,
we discuss its representations.
The example of \g=$su(3)$ is studied in full in section~\ref{Csu3}
where we construct the general $su(3)$ states and show the $su(3)$ 
representations contained in
the $\Sigma_{m_s}=\Sigma_2,\Sigma_4$ irreducible multiplets.

\section{The standard BRST complex and harmonic states}
\label{SBchs}

Let \g\ be defined by
\begin{equation}
[X_i,X_j]={C_{ij}}^k X_k\quad,\quad i,j,k=1,\dots,r\equiv\text{dim}\,\g
\quad,
\end{equation}
where $\{X_i\}_{i=1}^r$ is a basis of \g.
For instance, we may think of $\{X_i\}$ as a basis for the 
left invariant (LI) vector fields $X_i^L (g)\equiv X_i(g)$ on
the group manifold $G$ ($X_i(g) \in\mathfrak{X}^L(G)$).

Let $V$ be a vector space. In the Chevalley-Eilenberg
formulation (CE)
\cite{Che.Eil:48} of 
the Lie algebra cohomology,   
the space of $q$--dimensional cochains
$C^q(\g,V)$ is spanned by the $V$--valued skew--symmetric mappings
\begin{equation}
\psi: \g \wedge \mathop{\cdots}\limits^q \wedge \g \to V \quad,
\quad
\psi(g) = {1\over q!} \psi^A_{i_1\dots i_q} 
\omega^{i_1}(g)\wedge\dots\wedge
\omega^{i_q}(g) \otimes e_A
\quad,
\label{2.2}
\end{equation}
where the $\{\omega^i(g)\}$ form a basis of $\g^*$ (LI one-forms on $G$),
dual to the basis $\{X_i\}$ of LI vector fields on $G$, 
and the index $A=1,\dots,\text{dim}\,V$ 
labels the components in $V$.
Let $\rho$ be a representation of \g\ on $V$ 
($\rho:\g\to \text{End}(V)$).
The action of the Lie algebra coboundary operator 
$s_\rho$, ${s_\rho}^2=0$,
on the $q$--cochains $\psi^A$ (\ref{psigen}) is given by

\begin{definition}
[{\normalfont\textit{Coboundary operator}}{}]
The coboundary operator
$s_\rho: C^{q} (\g , V) \to C^{q{+}1} (\g, V )$
is defined by
\begin{equation}
\begin{array}{r@{}l}
\displaystyle
( s_\rho  \psi )^A \, ( X_{1} ,..., X_{q{+}1} ) & :=
\displaystyle
\sum_{i{=}1}^{q{+}1}
(-1)^{i + 1} \rho ( X_i )^A_{.B} \,( \psi^B ( X_{1} ,..., {\hat X}_{i},...,
X_{q{+}1} ) )
\\[0.3cm]
 &
\displaystyle
+ \sum_{{j,k=1 \atop j < k}}^{q{+}1} (-1)^{j{+}k} \psi^A ( [
X_{j} , X_{k} ] , \ X_{1} ,..., {\hat X}_{j} ,...,{\hat X}_{k},..., X_{q{+}1} )
\quad.
\end{array}
\label{2.3}
\end{equation}
\end{definition}
The space of $q$--\emph{cocycles} $Z_\rho^q(\g,V)$ 
(\emph{i.e.} $\text{Ker}\,s$) modulo the
$q$--coboundaries $B_\rho^q(\g,V)$ (\emph{i.e.} $\text{Im}\,s$) defines the
$q$--th Lie algebra cohomology group $H_\rho^q(\g,V)$.

Since we are assuming  $\g$ semisimple,
Whitehead's lemma states that, for
$\rho$ non-trivial, 
\begin{equation}
\label{whitelem}
H_\rho^q(\g,V)=0\quad,\quad \forall q\ge 0\quad,
\end{equation}
and we can restrict ourselves to $\rho=0$ cohomology for 
which the action of $s_\rho$ reduces to the second term in the r.h.s. of
eqn. (\ref{2.3}).

For the trivial representation, $s$ acts on $\psi$ (\ref{BRSTCOC}) in 
the same manner as the exterior derivative $d$ acts on LI forms.
It is then clear that we may replace the $\{ \omega^i (g)\}$ by the
ghost variables $\{ c^i \}$, 
\begin{equation}
c^i c^j = - c^j c^i \quad (\{c^i,c^j\}=0 \, , \quad
 \{c^i,{\partial \over \partial c^j}\} = \delta^i_j)
\quad,\quad i,j=1,\dots,r \quad,
\end{equation}
and the space of $q$--cochains by 
polynomials of (ghost) number $q \leq r$. 
The BRST operator~(\ref{exprs}) $s=s_2$ 
(the subindex 2 is added for convenience; its meaning will 
become clear in section~\ref{HoBo})
may be taken as the coboundary operator for the
($\rho$=0) Lie algebra cohomology \cite{Bon.Cot:83}. Indeed, the relations 
\begin{equation}
s_2 c^k = -\frac{1}{2} {C_{ij}}^{k} c^i c^j
\quad\quad
(\text{or}\ s_2 c= -\frac{1}{2} [c,c]\;,\; c=c^i \rho(X_i)\,) \quad,
\label{stck}
\end{equation}
reproduce the MC equations. As a result,
the Lie algebra 
cohomology may be equivalently formulated in terms 
of skew-symmetric tensors on \g, LI forms on $G$, or polynomials in ghost 
space (see \eg\ \cite{Azc.Izq:95}).

In the sequel we shall introduce the Grassmann variables $\pi_i$ to 
refer to the `partial derivative' ${\partial / \partial c^i}$,
appropriate for using the `ghost representation' for 
the cochains/states. These two sets of variables ($c^i$, 
$\pi_j$) span a Clifford--like 
algebra\footnote{
\label{foot4}
The algebra (\ref{2.11}) can be represented by a Clifford algebra (see
\emph{e.g.} \cite{Hol:90}). Namely, if we define
$\displaystyle c_i={1\over 2}(\gamma_i - i \gamma_{i+r}),\
\pi_i = {1\over 2}(\gamma_i + i \gamma_{i+r}),\ i=1,\dots, r$, where the
$\gamma$'s are the generators of a $2r$--dimensional 
Clifford algebra, then
$c_i$ and $\pi_i$ verify the relations (\ref{2.11}).}
defined by
\begin{equation}
\{c_i,\pi_j\} = \delta_{ij}\quad,\quad
\{c_i,c_j\}=0=\{\pi_i,\pi_j\}\quad.
\label{2.11}
\end{equation}
The algebra (\ref{2.11}) admits the (order reversing) involution
$\overline{\cdot}:c_i \mapsto \overline{c}_i = \pi_i$,
$\pi_i \mapsto \overline{\pi}_i= c_i$.
The \emph{anti}--BRST \emph{operator} $\overline{s}_2$ is given by
\begin{equation}
\overline{\cdot}: s_2 \mapsto \overline{s}_2 \, \, ,
\quad \quad \quad \quad \quad 
\overline{s}_2 = 
{1\over 2} {C_{ij}}^k c_k \pi^i \pi^j \quad
\left( =  {1\over 2} {C_{ij}}^k c_k
{\partial\over \partial c_i} {\partial\over \partial c_j} \right)
\, ,
\label{2.12}
\end{equation}
and it is also nilpotent. Denoting the space of the
BRST $q$--cochains (\ref{BRSTCOC}) by $C^q(\g)$, it follows that
\begin{equation}
s_2: C^q(\g) \to C^{q+1}(\g)
\quad,\quad
\overline{s}_2: C^{q}(\g) \to C^{q-1}(\g)
\quad.
\end{equation}

The presence of a metric ($\delta_{ij}$) on \g\ allows us to 
introduce the $*$--operator ($*: C^q(\g) \to C^{r-q}(\g)$) 
in the standard way. On $q$--forms on $G$,
\begin{equation}
(*\psi) = {1\over q!} {1\over (r-q)!}
\epsilon_{i_1 \dots i_r} \psi^{i_{1}\dots i_q}
\omega^{i_{q+1}}\wedge \dots \wedge\omega^{i_{r}}
\quad,
\label{2.17}
\end{equation}
and
\begin{equation}
 *^2 = (-1)^{q(r-q)} = (-1)^{q(r-1)} \, . 
\label{starsq}
\end{equation}
The scalar product of two LI $q$--forms on $G$, 
$\langle \cdot,\, \cdot \rangle : 
C^q(\g) \otimes C^q(\g) \to \mathbb{R}$ 
is then given by
\begin{equation}
\begin{array}{r@{}l}
\langle \psi',\psi \rangle :=
&
\displaystyle
\int_G \psi'\wedge *\psi
\\[0.35cm]
=
&
\displaystyle
\int_G {1\over q!^2} {1\over (r-q)!}
\psi'_{j_1\dots j_q}
\epsilon_{i_1 \dots i_{q} j_{q+1} \dots j_{r}}
\psi^{i_{1}\dots i_q}
\epsilon^{j_1 \dots j_r}
\omega^1 \wedge \dots \wedge \omega^r
\\[0.35cm]
=
&
\displaystyle
\int_G {1\over q!^2}  \epsilon^{j_1\dots j_q}_{i_1\dots i_q}
\psi'_{j_1\dots j_q} \psi^{i_1 \dots i_q} 
\omega^1 \wedge \dots \wedge \omega^r
=
{1\over q!} \psi'_{j_1\dots j_q} \psi^{j_1 \dots j_q}
\int_G \omega^1 \wedge \dots \wedge \omega^r
\end{array}
\label{2.18}
\end{equation}
and, normalising  the (compact) group volume
$\int_G \omega^1 \wedge \dots \wedge \omega^r$  to 1,  
reduces to~(\ref{scalprod}).
Clearly\footnote{
Using the $c$'s to write $\psi$ (eqn.~(\ref{BRSTCOC})), rather
than the $\omega$'s of~(\ref{2.18}), one might introduce a
Berezin~\cite{Ber:79b} integral measure to define
$\langle \psi',\psi \rangle$
above as $\int dc^1 \ldots dc^r {\psi'}^\dagger \psi$~\cite{Hol:90}
for states $\psi'$ and $\psi$ of ghost numbers $q$ and $r-q$ respectively.
However, this leads to a product which is not positive definite
\cite{Hol:90} and, moreover, does not have the natural geometrical
interpretation above.}

\begin{equation}
\langle \psi',\psi \rangle = \langle \psi, \psi' \rangle
\quad,\quad
\langle \psi,\psi \rangle > 0 \quad \forall \psi \neq 0
\quad.
\label{2.19}
\end{equation}

The codifferential $\delta$ is introduced, as usual, as the adjoint
of the
exterior derivative $d$, \emph{i.e.}, for a $(q-1)$--form $\psi'$,
\begin{equation}
\begin{array}{r@{}l}
\langle d \psi',\psi \rangle =
&
\displaystyle
\int_G d\psi' \wedge * \psi =
(-1)^q \int_G \psi' \wedge d*\psi =
(-1)^{q + (q-1)(r-q+1)} \int_G \psi' \wedge * (* d * \psi)
\\[0.35cm]
\equiv
&
\displaystyle
\int_G \psi' \wedge * \delta \psi
=\langle \psi', \delta \psi \rangle
\quad,
\end{array}
\end{equation}
so that 
\begin{equation}
\delta=(-1)^{r(q+1)+1} * d *
\quad, \quad
(d = (-1)^{r(q+1)} * \delta *) \quad,\quad 
\delta^2=0
\, \, .
\label{2.22}
\end{equation}

The correspondence $\omega^i(g) \leftrightarrow c^i$, 
$d \leftrightarrow s_2$
above allows us to translate all this into the BRST language.
First one checks, on any BRST $q$--cochain~(\ref{BRSTCOC}), that the
basic operators $c^i$ and $\pi^i$ are transformed by $*$ according to
\begin{equation}
\pi^i = (-1)^{r(q+1)} * c^i *
\quad,\quad
c^i = (-1)^{r(q+1)+1} * \pi^i *
\quad,
\label{2.23}
\end{equation}
so that
\begin{equation}
\begin{array}{c}
*(c^{i_1}\dots c^{i_{2k}}) * = (-1)^{(r+1)q+k}\pi^{i_1}\dots \pi^{i_{2k}}
\ , \ 
*(c^{i_1}\dots c^{i_{2k+1}}) * = (-1)^{r(q+1)+k}\pi^{i_1}\dots \pi^{i_{2k+1}}
\ ,
\\[0.3cm]
*(\pi^{i_1}\dots \pi^{i_{2k}}) * = (-1)^{(r+1)q+k}c^{i_1}\dots c^{i_{2k}}
\ ,\ 
*(\pi^{i_1}\dots \pi^{i_{2k+1}}) * = (-1)^{r(q+1)+k+1}c^{i_1}\dots c^{i_{2k+1}}
\ .
\end{array}
\label{ssign}
\end{equation}

As a consequence of (\ref{2.23}) one finds for
$\psi' \in C^{q+1}(\mathcal{G})$,  $\psi \in C^q(\mathcal{G})$,
\begin{equation}
\begin{array}{r@{}l}
\langle \psi' , c^i \psi \rangle
= &
\displaystyle
\int_G \psi'\wedge * c^i \psi =
\displaystyle
(-1)^{q(r-q)}\int_G
\psi'\wedge * c^i * * \psi =
(-1)^q
\int_G
\psi'\wedge \pi^i * \psi
\\ [0.35cm]
= &
\displaystyle
\int_G \pi^i \psi'\wedge * \psi
=\langle\pi^i \psi' , \psi \rangle \quad,
\end{array}
\end{equation}
using the fact that $\pi^i$ is a graded derivative and that
$\psi'\wedge * \psi\equiv 0$. 
Thus, $c^i$ and $\pi^i$ are adjoints to each other
with respect to the inner product $\langle\ ,\ \rangle$
or, in other words, the involution $\overline{\cdot}$  
defines the adjoint with respect to $\langle\ ,\ \rangle$.
Thus, $s_2\sim d$ and (\ref{2.22}) lead to
\begin{equation}
 \overline{s}_2 = (-1)^{r(q+1)+1} * s_2 *
\label{brstadj}
\end{equation}
since
\begin{equation}
\begin{array}{r@{}l}
\delta
&
\displaystyle
=
(-1)^{r(q+1)+1} * d *
\sim (-1)^{r(q+1)+1} * s_2 *
= - (-1)^{r(q+1)+1} {1\over 2} {C_{ij}}^k  * c^i c^j \pi_k *
\\[0.3cm]
&
\displaystyle
= - (-1)^{r(q+1)+1 + (q-1)(r-q+1) + q(r-q)}
{1\over 2} {C_{ij}}^k  * c^i * * c^j * * \pi_k *
=
{1\over 2} {C_{ij}}^k   \pi^i   \pi^j  c_k = \overline{s}_2
\; .
\end{array}
\end{equation}

{}The anticommutator of the nilpotent
 operators $s_2$ and $\overline{s}_2$ defines the Laplacian 
$\Delta \equiv  W_2\,$, $W_2: 
C^q(\g) \to C^q(\g)$, 
\begin{equation}
W_2:=\{s_2,\overline{s}_2\} = ( s_2 + \overline{s}_2 ) ^2
\quad.
\label{2.26}
\end{equation}
The operators $W_2,\ s_2,\ \overline{s}_2$ generate the
supersymmetry algebra $\Sigma_2$
\begin{equation}
[ s_2, W_2 ] = 0
\quad,\quad
[\overline{s}_2, W_2]=0
\quad,\quad
\{s_2,\overline{s}_2\}=W_2
\quad.
\label{2.27}
\end{equation}
$\Sigma_2$ has the structure of a central extension of ($s_2$,
$\overline{s}_2$) by $W_2$, the Laplacian being the central generator.
The operator $W_2$ is invariant under the involution $\overline{\cdot}$
($W_2=\overline{W}_2$) and commutes with $*$, since
\begin{equation}
\begin{array}{r@{}l}
* W_2 * & = * ( s_2 \overline{s}_2 + \overline{s}_2 s_2 ) * =
(-1)^{(q-1)(r-q+1)} * ( s_2 * * \overline{s}_2 + 
\overline{s}_2 * * s_2 ) *
\\[0.3cm]
&
=(-1)^{(q-1)(r-q+1) + r(q-1) + 1 + rq}
( \overline{s}_2 s_2 + s_2 \overline{s}_2 )
=(-1)^{q(r-q)} W_2
\quad,
\end{array}
\end{equation}
which, with the help of~(\ref{starsq}), implies $[W_2, \, *]=0$.
Then, as in the standard Hodge theory on compact Riemannian
manifolds, we have

\begin{lemma}
\label{lem2.1}
A BRST cochain $\psi$ is $W_2$--harmonic, $W_2\psi =0$, iff
it is $s_2$ \emph{and} $\overline{s}_2$--closed.
\end{lemma}
\begin{proof}
It is clear that if $s_2\psi=0=\overline{s}_2\psi$, then $W_2\psi =0$.
Now, if $W_2\psi =0$,
\begin{equation}
\begin{array}{r@{}l}
0
= \langle \psi , W_2 \psi \rangle =
\langle \psi , (s_2 \overline{s}_2 + \overline{s}_2 s_2) \psi \rangle
= \langle \overline{s}_2 \psi ,\overline{s}_2 \psi\rangle
+ \langle s_2 \psi , s_2 \psi\rangle
\end{array}
\quad;
\end{equation}
from (\ref{2.19}) easily follows that both terms have to be zero 
and hence $s_2\psi=0=\overline{s}_2\psi$.
\end{proof}
\begin{theorem}
\label{th2.1}
Each BRST cochain $\psi$ admits the Hodge decomposition
\begin{equation}
\psi = s_2 \alpha + \overline{s}_2 \beta + \gamma
\quad,
\end{equation}
where $\gamma$ is $W_2$--harmonic
(the proof of theorem.~\ref{th3.1} below includes this case).
\end{theorem}

To find the algebraic meaning of $W_2$,
let us write the generators $X_i$ on ghost space as
\begin{equation}
X_i \equiv -{C_{ij}}^k c^j \pi_k \quad .
\label{Xreal}
\end{equation}
They act on BRST cochains in the same way
as the Lie derivatives with respect to
the LI vector fields on $G$ act on LI forms on $G$:
\begin{equation}
X_i c^k = - {C_{ij}}^k c^j \, \, , 
\end{equation}
(\emph{cf.} $L_{X_i} \omega ^k = -{C_{ij}}^k \omega ^j$, in which 
$X_i \in \mathfrak{X}^L(G)$ and $\omega \in \mathfrak{X}^{*L}(G)$). 
The $X_i$ in~(\ref{Xreal}) are in the adjoint representation 
of \g\ 
and satisfy $\overline{X}_{i} = - X_{i}$ and $*X_{i} = X_{i}*$.
\emph{Invariant states} are those for 
which $X_i\psi=0$, $i=1,\dots,r$.

In terms of $X_i$, the operators $s_2$ and $\overline{s}_2$ may 
be written as
\begin{equation}
s_2 = {1\over 2} c^i X_i \, \, , \quad \quad
\overline{s}_2 = - {1\over 2} \pi^j X_j
\, \, .
\label{ststbX}
\end{equation}
Using the fact that $c^i$ and $\pi^j$ transform in the adjoint 
representation,
\begin{equation}
X_k c^i = - {C_{kr}}^i c^r
\quad,\quad
X_k \pi^i = - {C_{kr}}^i \pi^r
\, ,
\label{XkciXkpii}
\end{equation}
it is easy to see 
that\footnote{
Taking advantage of the Cartan formalism by means of the equivalences 
$s_2\sim d$,
$\overline{s}_2\sim (-1/2)L_j i_j$ (where $i_j$ indicates the inner
product) and $X_j\sim L_j$, we may rapidly find
$W_2\sim(-1/2)[dL_j i_j+L_j i_j d]$=
$(-1/2)L_j[di_j+i_jd]=(-1/2)L_j L_j$.}

\begin{equation}
W_2 = - {1\over 2} \mathcal{C}^{(2)} = - {1\over 2} \delta^{ij} X_i X_j
\quad,
\label{2.32}
\end{equation}
\emph{i.e.} the Laplace--type operator is proportional to 
the second order Casimir operator of the algebra.
\begin{remark}
The expression for $W_2$ in~\cite{Ger:86,Bar.Yan:87,Hol:90}
contains more terms due to the fact that these authors consider
$\rho \neq 0$ in general. But, as noticed in \cite{Hol:90}, $\rho=0$ is
the only possibility if we restrict ourselves to \emph{non--trivial}
harmonic states.
In fact, we prove here that this is a direct consequence
of Whitehead's lemma (\ref{whitelem}). Let $\tau$ be the operator
defined by its action on 
($V$--valued) $q$--cochains $\psi$ through
\begin{equation}
( {\tau} \psi )_{i_1 \dots i_{q-1}}^A =
k^{ij} \rho(X_i)^A_{B} \psi_{j i_1 \dots i_{q-1}}^B \quad.
\end{equation}
It may be verified that
\begin{equation}
[(s_\rho {\tau} + {\tau} s_\rho) \psi] ^A _{i_1 \dots i_q} =
\psi_{i_1 \dots i_q} ^B \mathcal{C}^{(2)}(\rho)^A_{B} \quad,
\end{equation}
where
$\mathcal{C}^{(2)}(\rho)^A_{B}
\equiv k^{ij} \rho(X_i)^A_{C} \rho(X_j)^C_{B}$ is the
Casimir operator for the representation $\rho$, and hence
proportional to $\delta^A_B$.
It then follows that for any $\rho\not=$0 $q$--cocycle $\psi$
($s_\rho \psi=0$)
\begin{equation}
s_\rho ( {\tau} \psi \mathcal{C}^{(2)}(\rho)^{-1} ) \propto \psi
\end{equation}
\emph{i.e.}, $\psi$ is a (trivially harmonic state)
coboundary generated by a
$(q-1)$--cochain proportional to $\tau \psi \mathcal{C}^{(2)}(\rho)^{-1}$, 
\emph{q.e.d.}
Hence, any non--trivial BRST--invariant state $\psi$ ($s_2\psi=0,\ \psi\neq 
s_2\varphi$) is a \g\ singlet and, as a consequence of Th.~\ref{th2.1}, its 
class contains a unique $W_2$ harmonic representative.
\end{remark}

{}From (\ref{2.32}) we also deduce the following
\begin{lemma}
A state $\psi$ is $W_2$--harmonic iff it is invariant\footnote{
The CE analogues to the BRST $q$--cochains, the LI $q$--forms on $G$,
automatically satisfy $L_{X^R} \psi =0$, since the RI vector
fields $X^R$ on $G$ generate the \emph{left} transformations. The
invariance under the right transformations ($L_X \psi=0$
where $X$ is a LI vector field, or $X \psi=0$ in the 
BRST formulation) is an \emph{additional} condition. 
Thus, invariance above really means \emph{bi-invariance}
(under the left and right group translations) in the CE formulation of Lie
algebra cohomology.}, $X_i \psi =0$.
\label{W2harminv}
\end{lemma}
\begin{proof}
If $\psi$ is invariant, $W_2\psi = \displaystyle
-{1\over 2} \delta^{ij} X_i X_j \psi =0$.
If $\psi$ is $W_2$--harmonic,
\begin{equation}
0 = \langle \psi, W_2\psi\rangle =
- {1\over 2}\langle \psi, \delta^{ij} X_i X_j \psi\rangle =
{1\over 2} \delta^{ij} \langle X_i\psi,  X_j \psi\rangle
\end{equation}
and $X_j \psi=0$, since $\langle\ ,\ \rangle$ is non--degenerate, \emph{q.e.d.}
In fact, if $\psi$ is invariant, $\psi$ is both $s_2$ and $\overline{s}_2$ 
closed by (\ref{ststbX}).
\end{proof}
\begin{corollary}
Each non--trivial element in the cohomology ring $H^*(\g)$
may be represented by an invariant state.
\end{corollary}
\begin{proof}
Let $\psi\in Z(\g)$ be nontrivial.
Hence its decomposition has the form
\begin{equation}
\psi = s_2 \alpha + \gamma \quad.
\end{equation}
Therefore $\psi -s_2\alpha$ is in the cohomology class of $\psi$ and
is harmonic (and hence invariant).
\end{proof}

\section{Higher order BRST and anti--BRST operators}
\label{HoBo}
\subsection{Invariant tensors}
The considerations of the previous section rely on the
nilpotent operator $s_2$ and its adjoint, both 
constructed out of the structure constants $C_{ijk}$.
The latter determine a skew-symmetric tensor of order three which
can be seen as a third order cocycle $C=C_{ijk} c^i c^j c^k$ and, 
additionally, is invariant under the action of the Lie algebra
generators $X_k$.
Indeed, acting on $C$ with the $X$'s one gets a sum
of three terms, in each of which one of the indices of $C$ is
transformed in the adjoint representation and the statement of
invariance is equivalent to the Jacobi identity.
Notice that we need not
saturate every index of $C_{ijk}$ with the same type of variable in order
to get an invariant quantity  ---it suffices that each type of 
variable transforms in the adjoint
representation 
(for example, $s_2$ in (\ref{exprs}), which
is also invariant, involves saturating two $c$'s and one $\pi$).

The cohomology of simple Lie
algebras contains, besides the three cocycle $C$ above, other, higher
order skewsymmetric tensors with similar properties. As mentioned
in the introduction, any compact simple Lie algebra $\g$ of rank $l$ has
$l$ primitive cocycles given by skew-symmetric tensors
$\Omega^{(2m_s-1)}_{i_1  i_2 \ldots i_{2m_s-1}}$ ($s=1, \ldots , l$),
associated to the $l$
Casimir--Racah primitive invariants of rank $m_s$
\cite{Rac:50,Gel:50,Per.Pop:68,Azc.Bue:97}.
Their invariance is expressed by the equation
\begin{equation}
\sum_{j=1}^{2m_s-1}{C_{b i_j}}^{a} \Omega^{(2m_s-1)}_{i_1 i_2
\ldots \hat{i}_j a i_{j+1} \ldots i_{2m_s-1}} = 0 \quad .
\label{Omegainv}
\end{equation}
Due to the MC equations, the above relation implies that
$\Omega ^{(2m_s-1)}_{i_1 \ldots i_{2m_s-1}} \omega^{i_1}  
\ldots \omega^{i_{2m_s-1}}$ is a 
cocycle for the Lie algebra coboundary operator
(\ref{2.3}) for $\rho$=0 (in
the language of forms, this is equivalent to saying that any
bi--invariant form is closed, and hence a CE cocycle). The 
existence of these
cocycles is related to the topology of the corresponding group
manifold, in particular to the odd-sphere product structure
that the simple compact group manifolds have from the point of view
of real homology  
(see \emph{e.g.}~\cite{Che.Eil:48,Hod:41,Sam:52,Gre.Hal.Van:76,Azc.Izq:95}). 
We may use the correspondence $c^i
\leftrightarrow \omega^i$ and the discussion after
theorem~\ref{th2.1} to move freely from the CE approach to the BRST
one here.

Let us consider for definiteness the case of $su(n)$,  
for which $m_1=2$, $m_2=3, \dots , m_l=n$
and there exist $l$=$(n-1)$ different primitive
skew--symmetric tensors of rank $3, 5,
\dots , 2n-1$. Consider, for a given $m$, the $(2m-1)$--form
\begin{equation}
\Omega ^{(2m-1)} = \frac{1}{(2m-1)!} \mbox{Tr} (\theta \wedge
\mathop{\cdots}\limits ^{2m-1} \wedge \theta )
\, ,
\label{cocycledef}
\end{equation}
where $\theta \equiv \omega ^i T_i$ and $T_i\in\g$ is in the
defining representation of $su(n)$. Since $d
\Omega ^{(2m-1)}=0$, the coordinates of $\Omega ^{(2m-1)}$ 
provide a $(2m-1)$--cocycle on $su(n)$. One can show 
(see \emph{e.g.} \cite{Azc.Mac.Mou.Bue:97}) that
\begin{equation}
\Omega ^{(2m-1)}_{\rho i_2 \ldots i_{2m-2} \sigma}
=
{1\over  (2m-3)!}
k_{\rho l_1 \ldots l_{m-1}} {C_{j_2 j_3}} ^{l_1} \ldots {C_{j_{2m-2}
\sigma}} ^{l_{m-1}} \epsilon ^{j_2 
\ldots j_{2m-2}}_{i_2 \ldots i_{2m-2}}
\, \, ,
\label{Omegacoord}
\end{equation}
is a skew--symmetric tensor,
where
\begin{equation}
k_{\rho l_1 \ldots l_{m-1}} = \mbox{sTr} (T_{\rho} T_{l_1} \ldots
T_{l_{m-1}})
\label{kcoord}
\end{equation}
is a symmetric invariant tensor given by the 
symmetrised trace of a product of $m$
generators (its invariance can be expressed by an equation similar
to~(\ref{Omegainv})).
Symmetric invariant tensors $k_{i_1 \dots i_m}$ give rise to
Casimir-Racah operators 
\begin{equation}
\C^{(m)} = k ^{i_1 \ldots i_{m}} X_{i_1} \ldots X_{i_{m}} 
\label{Cmdef}
\end{equation}
which commute with the generators; $\C^{(2)}$ is the 
standard quadratic Casimir operator.

\subsection{Higher order operators}
The above family of
cocycles $\Omega$ can be used to construct \emph{higher--order} 
BRST \emph{operators} \cite{Azc.Bue:97,Azc.Izq.Per:98}. 
To each invariant tensor of rank $m_s$ corresponds 
a BRST operator $s_{2m_s-2}$ which, in terms of the coordinates 
${\Omega_{i_1 \ldots i_{2m_s-2}}}^\sigma$ of the $(2m_s-1)$--cocycle 
(\ref{Omegacoord}), is given by
\begin{equation}
s_{2m_s-2} = -\frac{1}{(2m_s-2)!} 
{\Omega^{(2m_s-1)}_{i_1 \ldots i_{2m_s-2}}}^\sigma  
c^{i_1} c^{i_2} \ldots c^{i_{2m_s-2}} \pi_{\sigma}
\, \, .
\label{smjdef}
\end{equation}
These operators are particularly interesting in view of the property
\begin{equation}
\{ s_{2m_s-2}, \ s_{2m_{s'}-2} \} = 0\quad,\quad
s,s'=1,\dots, l \quad ;
\label{ssanticom}
\end{equation}
\emph{i.e.}, they are nilpotent and anticommute (see 
\cite{Azc.Bue:97} for a proof). 

For each $m_s$, $s>1$, we may look at
$s_{2m-2}$ \footnote{
We shall often write $m$ for $m_s$ henceforth.}
as a \emph{higher order coboundary} operator, 
$s_{2m-2}:\, C^q(\g)\to C^{q+(2m-3)}(\g)$. The analogue
of the MC equation (\ref{stck}) for $s_{2m-2}$ is given by
\begin{equation}
s_{2m-2} c^a = 
-\frac{1}{(2m-2)!} 
{\Omega^{(2m-1)}_{i_1 \ldots i_{2m-2}}}^a  
c^{i_1} c^{i_2} \ldots c^{i_{2m-2}}
\label{genmc}
\end{equation}
which may also be written as
\begin{equation}
s_{2m-2} c = - {1\over (2m-2)!} [c,\mathop{\cdots}\limits^{2m-2},c]\quad,
\label{shomc}
\end{equation}
where $[c,\mathop{\cdots}\limits^{2m-2},c]:=c^{i_1}\dots c^{i_{2m_2}} 
[T_{i_1},\dots,T_{i_{2m-2}}]$ and the higher--order structure constants of the 
$(2m-2)$--bracket \cite{Azc.Bue:97} are given by the $(2m-1)$ cocycle, \ie
\begin{equation}
[T_{i_1},\dots,T_{i_{2m-2}}]= {\Omega^{(2m-1)}_{i_1 \ldots i_{2m-2}}}^a T_a
\quad.
\label{multiT}
\end{equation}
Using (\ref{shomc}), the nilpotency of $s_{2m-2}$ follows from the higher 
order Jacobi identity
\begin{equation}
s^2_{2m_2} c =  - {1\over (2m-2)!} {1\over (2m-3)!}
[c,\mathop{\cdots}\limits^{2m-3},c, [c,\mathop{\cdots}\limits^{2m-2},c]]=0
\label{SGJI}
\end{equation}
which the r.h.s. of (\ref{SGJI}) satisfies as a consequence of 
${\Omega^{(2m-1)}_{i_1 \ldots i_{2m-2}}}^a$ being a cocycle.

Moreover, for each \g\ we may introduce the \emph{complete 
BRST operator} $\mathbf{s}$ \cite{Azc.Bue:97}
\begin{equation}
\begin{array}{r@{}l}
\mathbf{s}=
\displaystyle
&
\displaystyle
-{1\over 2}{C_{j_1j_2}}^\sigma c^{j_1}c^{j_2} \pi_\sigma
-\ldots-
{1\over (2m_s-2)!}{\Omega^{(2m_s-1)}_{j_1\ldots j_{2m_s-2}}}^\sigma
c^{j_1}\ldots c^{j_{2m_s-2}} \pi_\sigma
-\ldots
\\[0.3cm]
&
\displaystyle
-
{1\over (2m_l-2)!}{\Omega^{(2m_l-1)}_{j_1\ldots j_{2m_l-2}}}^\sigma
c^{j_1}\ldots c^{j_{2m_l-2}} \pi_\sigma
\equiv s_2+\ldots+s_{2m_s-2}+\ldots +s_{2m_l-2}
\end{array}
\label{scomplete}
\end{equation}
This operator is nilpotent, and its terms have (except for some 
additional
ones that break the generalised Jacobi identities which are at the
core of the nilpotency of $s_{2m_{s}-2}$) the same structure
as those which appear in closed string theory \cite{Wit.Zwi:92}
and lead to a strongly homotopy algebra \cite{Lad.Sta:93}.
In fact, the higher order structure constants (which here have definite 
values and 
a geometrical meaning as higher order cocycles of \g) correspond to the 
string correlation functions giving the string couplings.
Since the expression for $\mathbf{s}$ in the homotopy Lie algebra that 
underlies closed string theory already includes a term of the form
$f_{j_1} ^\sigma c^{j_1} \pi_\sigma$, $f$ nilpotent,
$\mathbf{s}^2=0$ is not satisfied (as it is for (\ref{scomplete}))
by means of a sum of independently satisfied Jacobi identities, and in
particular the ${C_{ij}}^k$ do not satisfy the Jacobi identity and
hence do not define a Lie algebra. 

For each $\smj$ we now introduce its 
adjoint \emph{anti}--BRST \emph{operator} $\smjb$,
\begin{eqnarray}
\smjb & = & - \tmjfi \Omis c_{\sigma} 
            \pi^{i_{2m-2}} \ldots \pi^{i_1} \nonumber \\
      & = & - \frac{(-1)^{m-1}}{(2m-2)!} 
            \Omis c_{\sigma} \pi^{i_1} \ldots
            \pi^{i_{2m-2}} \, . 
\label{smjbdef}
\end{eqnarray}
Each pair ($\smj$, $\smjb$)
allows us to construct a \emph{higher-order 
Laplacian} $\Wmj$ 
\begin{equation}
\Wmj  = (\smj + \smjb)^2 = \smj \smjb + \smjb \smj \quad. 
\label{Wmjdef}
\end{equation}
Clearly, $\smj$, $\smjb$ and $\Wmj$ all commute with the generators
$X_i$, and we have the following
\begin{lemma}
For each $s=1,\dots,l$, the higher order BRST and anti-BRST 
operators $s_{2m_s-2}$ and $\overline{s}_{2m_s-2}$,
together with their associated Laplacian $W_{2m_s-2}$ define the
superalgebra $\Sigma_{m_s}$
\begin{equation}
[s_{2m_s-2},W_{2m_s-2}]=0 \quad , \quad
[\overline{s}_{2m_s-2},W_{2m_s-2}]=0 \quad , \quad
\{s_{2m_s-2},\overline{s}_{2m_s-2}\}=W_{2m_s-2} \quad,
\label{ssbWalg}
\end{equation}
which has a central extension structure. 

For $s=1$, $m_1=2$, $W_2=\Delta$ and the above expressions
reproduce~(\ref{exprs}), (\ref{stck}), (\ref{2.12}) and~(\ref{2.27}).
\end{lemma}

The BRST (anti-BRST) operator
$\smj$ ($\smjb$), acting on a monomial in the $c$'s, 
raises (lowers) its ghost number by $2m -3$ 
while $\Wmj$ leaves the ghost number invariant and is self-adjoint.
We notice that all terms in $\smj$ ($\smjb$) contain one
($2m-2$) $\pi$, and that the term with the maximum number
of $\pi$'s in $\Wmj$ contains
(at most) $2m-2$ of them.
This is so because
the two terms with $2m-1$ $\pi$'s (from $\smj \smjb$, 
$\smjb \smj$) cancel, as one can  verify. The BRST operator
$\smj$ annihilates all states of  
ghost number $q > r-2m+2$ (since the product
of more than $r$ $c$'s necessarily vanishes) as well as zeroth
order states. Similarly, $\smjb$ annihilates states of 
ghost number $q<2m-2$
and the top state $c_1 \ldots c_r$. It follows that zero and
$r$--ghost number states are both $\Wmj$--harmonic. 

Let us establish now the relation between the 
$\overline{\cdot}$ operation
(the adjoint with respect to the inner 
product in (\ref{2.18}))
and the conjugation by the Hodge $*$ operator, as these apply to
$\smj$. 
\begin{lemma}
\label{lsdual}
The following equalities hold on any  state (BRST-cochain) 
of ghost number $q$,
\begin{equation}
\smjb= (-1)^{r(q+1)+1} *\smj*\,,\,\,
\smj = (-1)^{r(q+1)} * \smjb *\,,\,\,
\smj * = (-1)^q * \smjb \, \, ;
\label{ssmjs}
\end{equation}
\end{lemma}
notice that the sign factors do not depend on $m$ and hence 
they coincide with those of (\ref{2.22}). The proof is
straightforward, using (\ref{starsq}), (\ref{ssign}), where care 
should be taken to
substitute the ghost numbers actually `seen' by the operators. 

Although we shall not use them here we also introduce, for the sake
of completeness, the $V$--valued higher
order coboundary operators ${s_{\rho}}_{(2m_s-2)}$
for a non-trivial representation $\rho\in\text{End}\,V$
of the $(2m_s-2)$-algebra.
They are given by 
\begin{equation}
{s_{\rho}}_{(2m_s-2)} = 
c^{i_1}\dots c^{i_{2m_s-3}} \rho(X_{i_1}) \dots \rho(X_{i_{2m_s-3}})
-\frac{1}{(2m_s-2)!} 
{\Omega^{(2m_s-1)}_{i_1 \ldots i_{2m_s-2}}}^\sigma  
c^{i_1} c^{i_2} \ldots c^{i_{2m_s-2}} \pi_{\sigma}
\quad.
\label{sconro}
\end{equation}
It may be seen that the nilpotency of ${s_{\rho}}_{(2m_s-2)}$
is guaranteed by the fact that the skewsymmetric
product of ($2m_s$--2) $\rho$'s, which defines the multibracket
($2m_s-2$)-algebra for an appropriate $\rho$
(eqn. (\ref{multiT}) with $T\to\rho$),
satisfies the corresponding generalised Jacobi identity as before.

\subsection{Higher order Hodge decomposition and representations of
$\Sigma_{m_s}$}
\label{RhoHd}

Let us now look at the irreducible representations 
of (\ref{ssbWalg}), which have the same structure as
the supersymmetry algebra. Since $\Wmj$ commutes with $\smj$,
$\smjb$, each multiplet of states will have a fixed $\Wmj$--eigenvalue.
Let us call $\gamma$ a $\Wmj$--\emph{harmonic state} iff
$\Wmj \gamma =0$. Then lemma~\ref{lem2.1}
transcribes trivially to the present higher order case
so that $\gamma$ is harmonic iff it is
$\smj$ and $\smjb$--closed. Hence a
harmonic state $\gamma$ is a singlet of $\Sigma_m$.
We may also extend theorem~\ref{th2.1} and prove the following
\begin{theorem}[{\normalfont\textit{Higher order Hodge
decomposition
}}{}]
\label{th3.1}
Each BRST cochain $\psi$ admits a unique decomposition 
\begin{equation}
\psi = \smj \alpha + \smjb \beta + \gamma
\end{equation}
where $\gamma$ is $\Wmj$--harmonic.
\end{theorem}
\begin{proof}
We denote by $\S$ the space of all states (\emph{i.e.} skewsymmetric 
polynomials in the $c$'s), $\Km$
the kernel of $\Wmj$ ($\Wmj$--harmonic space) 
and $\Km^{\perp}$ the complement of $\Km$ in
$\S$. Let $\Pzw$ be the projector from $\S$ to $\Km$.
Let $\psi\in\S$; then, $(1-\Pzw)\psi$ lies in $\Km^\perp$. 
However, since the restriction of $\Wmj$ to $\Km^\perp$ is invertible,
there exists a unique $\phi$ in $\Km^\perp$ such that 
$(1-\Pzw)\psi=\Wmj \phi$, from which we get
\begin{eqnarray}
\psi & = & \Wmj \phi +\Pzw \psi \nonumber \\
       & = & \smj (\smjb \phi) + \smjb (\smj \phi) + \Pzw \psi
\quad ,
\label{Hodgedec}
\end{eqnarray}
which provides the desired
decomposition of $\psi$ with $\alpha=\smjb \phi$,\, $\beta=\smj \phi$\,
and $\gamma= \Pzw \psi$, \emph{q.e.d.}
\end{proof}

To complete the analysis of the irreducible representations of $\Sigma$
consider now an eigenstate $\chi$ of $\Wmj$
for non-zero (and hence positive) eigenvalue $w$,
$\Wmj \chi = w \chi$, $w > 0$. This gives rise to the states 
\begin{equation}
\phi \equiv \smj \chi \quad , \quad \quad \quad \quad
\psi \equiv \smjb \chi \quad , \quad \quad \quad \quad
\sigma \equiv \smj \smjb \chi \quad .
\label{phipsisidef}
\end{equation}
 Further applications of $\smj$ or
$\smjb$ produce linear combinations of the above states, for
example $\smjb \smj \chi = \Wmj \chi - \smj \smjb \chi = w \chi -
\sigma$ \emph{etc}. The quartet $\{\chi, \, \phi, \, \psi, \, 
\sigma \}$ collapses to a doublet if
either $\smj \chi =0$ or $\smjb \chi =0$. In this case, 
$\chi$ is the Clifford vacuum and  $\smj$,
or $\smjb$, respectively, plays the role of the annihilation operator.
Let $\chi$ be neither $\smj$ nor
$\smjb$-closed. The state $\sigma$ of (\ref{phipsisidef}) is, by
construction, $\smj$-closed. Then, we can always choose a linear
combination of $\chi$ and $\sigma$ that is $\smjb$-closed.
Indeed, 
for the $\{\chi, \, \phi, \, \psi, \, \sigma \}$ of
(\ref{phipsisidef}) we easily compute
\begin{eqnarray}
\Vert \phi \Vert ^2 + \Vert \psi \Vert ^2 & = & 
              \langle \smj \chi \, , \, \smj \chi \rangle 
             +\langle \smjb \chi \, , \, \smjb \chi \rangle \nonumber \\
 & = & \langle \chi \, , \, \Wmj \chi \rangle \quad = \quad  w \, \, ,
\end{eqnarray}
where we have taken $\Vert \chi \Vert ^2 \equiv 
\langle \chi \, , \, \chi \rangle = 1$. Setting 
\begin{equation}
q=\sqrt{w} \, , \quad \quad \quad 
\Vert \phi \Vert = q \sin \theta \, , \quad \quad \quad
\Vert \psi \Vert = q \cos \theta \, ,
\label{qthetadef}
\end{equation}
we find that the following linear combinations
\begin{equation}
\chi' = \frac{1}{q \sin \theta} (q \chi - q^{-1} \sigma) \, ,
\quad \quad 
\sigma ' = \frac{1}{q^2 \cos \theta} \sigma \, ,
\quad \quad 
\phi ' = \frac{1}{q \sin \theta} \phi \, ,
\quad \quad 
\psi ' = \frac{1}{q \cos \theta} \psi \, ,
\label{primebas}
\end{equation}
form an orthonormal set, with the doublet $\{\chi' \, , \, \phi' \}$
 satisfying
\begin{equation}
\smjb \chi' =0 \quad,\quad
\smj \chi' = q \phi'\quad;\quad
\smj \phi' =0
\quad,\quad
\smjb \phi' = q \chi'
\quad,
\end{equation}
and similarly for $\{ \psi ' \, , \, \sigma ' \}$, \emph{i.e.} the two
doublets decouple.
Notice that $\theta$ in (\ref{qthetadef}), 
$0 < \theta < \pi/2$, is the angle between $\chi$ and $\sigma$ in the
$\chi$-$\sigma$ plane:
\begin{equation}
\langle \chi, \, \sigma \rangle = q^2 \cos^2\theta= \Vert \chi
\Vert \cdot \Vert \sigma \Vert \cos \theta \, \, . 
\label{chisiangle}
\end{equation}
Once in the primed basis of (\ref{primebas}), 
$\smj
\smjb \xi$ (where $\xi$ stands for any of the four primed states
above) is equal to either $w \xi$ or $0$ and hence, $\smj \smjb$
commutes with all operators that commute with $\Wmj$ (similarly for
$\smjb \smj$). 
Thus, the representation of the different superalgebras $\Sigma_m$ fall, in all 
cases, into singlets (harmonic states) and pairs of 
doublets.
Singlets and doublets here are the trivial analogues of the
`short' (massless) and `long' (massive) multiplets of the standard 
supersymmetry algebra.

\medskip

Owing to the particular importance of harmonicity, we
investigate the relation between the kernel of a higher-order 
Laplacian and that of $W_2 \propto \mathcal{C}^{(2)}$. To this end, we
rewrite $\smj$ as  (Greek indices below also
range in $1, \ldots , r \equiv \dim \g$).
\begin{eqnarray}
\smj & = & - \frac{1}{(2m-2)!} \Omis c^{i_1} \ldots c^{i_{2m-2}} 
                                            \pi_\sigma \nonumber \\
     & = & - \frac{1}{(2m-2)!} {k_{j_1 \ldots j_{m-1}}}^\sigma 
	      {C_{\rho i_2}}^{j_1}
              \ldots    {C_{i_{2m-3} i_{2m-2}}}^{j_{m-1}} c^\rho c^{i_2}
              \ldots c^{i_{2m-2}} \pi_\sigma \nonumber \\
     & = & \frac{1}{(2m-2)!} \Bigl(\sum_{r=2}^{m-1} {k_{i_2 j_2 \ldots 
              \widehat{j_r} \alpha
              \ldots j_{m-1}}}^\sigma {C_{\rho j_r}}^\alpha 
              {C_{i_3 i_4}}^{j_2} \ldots 
              {C_{i_{2r-1} i_{2r}}}^{j_r} \ldots 
              {C_{i_{2m-3} i_{2m-2}}}^{j_{m-1}} 
              \nonumber \\
     &   & + {k_{i_2 j_2 \ldots j_{m-1}}}^\alpha 
             {C_{\alpha \rho}}^\sigma
             {C_{i_3 i_4}}^{j_2} \ldots 
             {C_{i_{2m-3} i_{2m-2}}}^{j_{m-1}} \Bigr) 
	     c^\rho c^{i_2} \ldots c^{i_{2m-2}} \pi_\sigma \nonumber \\
     & = &  \frac{1}{(2m-2)!} 
	    {k_{\beta j_1 \ldots j_{m-2}}}^\alpha {C_{i_1 i_2}}^{j_1}
              \ldots {C_{i_{2m-5} i_{2m-4}}}^{j_{m-2}} 
	     c^\beta c^{i_1} \ldots c^{i_{2m-4}} X_\alpha \, \, ,
\label{brstX}
\end{eqnarray}
where the invariance of ${k_{j_1 \ldots j_{m-1}}}^\sigma$ has been
used in the third line and the Jacobi identity in the last
equality.
Similarly, $\overline{s}_{2m-2}$ may be written as
\begin{equation}
\overline{s}_{2m-2} = - (-1)^m\frac{1}{(2m-2)!} 
            {k_{\beta j_1 \ldots j_{m-2}}}^\alpha {C_{i_1 i_2}}^{j_1}
              \ldots {C_{i_{2m-5} i_{2m-4}}}^{j_{m-2}} 
             \pi^\beta \pi^{i_1} \ldots \pi^{i_{2m-4}} X_\alpha \, \, ,
\label{antiX}
\end{equation}
This proves the following
\begin{lemma}
\label{cuasilie}
The higher order BRST and anti--BRST operators $s_{2m-2}$, 
$\overline{s}_{2m-2}$ may be written as
\begin{equation}
\smj = \Omega ^\alpha X_\alpha
\quad,\quad
\overline{s}_{2m-2} = - \overline{\Omega}^\alpha X_\alpha 
\quad,
\label{smjOX}
\end{equation}
where
\begin{equation}
\Omega ^\alpha \equiv 
\frac{1}{(2m-2)!} {k_{\beta j_1 \dots j_{m-2}}}^\alpha {C_{i_1 i_2}}^{j_1}
     \ldots {C_{i_{2m-5} i_{2m-4}}}^{j_{m-2}} c^\beta c^{i_1} \ldots
     c^{i_{2m-4}} \quad,
\label{Omegaalpha}
\end{equation}
\begin{equation}
\overline{\Omega}^\alpha
\equiv 
(-1)^m \frac{1}{(2m-2)!} {k_{\beta j_1 \dots j_{m-2}}}^\alpha {C_{i_1 i_2}}^{j_1}
     \ldots {C_{i_{2m-5} i_{2m-4}}}^{j_{m-2}} \pi^\beta \pi^{i_1} \ldots
     \pi^{i_{2m-4}} \quad.
\label{antiOmegaalpha}
\end{equation}
\end{lemma}
For $m=2$ one gets $\displaystyle \Omega ^\alpha = {1\over 2} c ^\alpha$,
$\displaystyle \overline{\Omega} ^\alpha = {1\over 2} \pi ^\alpha$
and the expression~(\ref{ststbX}) for $s_2$, $\overline{s}_2$ is recovered.
As a $W_2$--harmonic state is invariant, the above relation shows that
the kernel of $W_2$ is contained in the kernel of $W_{2m_s-2}$ for all
$m_s$, $s=2, \ldots , l$.
It follows from (\ref{smjOX}) that an invariant state is
$s_{2m-2}$ and $\overline{s}_{2m-2}$ closed, and hence the following lemma
\begin{lemma}
Every invariant state is $W_{2m-2}$ harmonic.
\end{lemma}

In the particular realization of the $\Sigma_{m_s}$
algebra (\ref{ssbWalg}) in terms of
ghosts and antighosts given in (\ref{smjdef}), (\ref{smjbdef}),
$W_{2m_s-2}$ is ghost number preserving and commutes with the Lie algebra 
generators $X_i$.
There exists therefore a basis of the $c$'s in 
which $W_{2m_s-2}$, $s=1, \dots ,l$ 
is diagonal. For a fixed ghost number $q$,  the 
$\displaystyle \left({r \atop q}\right)$ 
independent monomials $c_{i_1} \ldots c_{i_q}$ transform as
the fully antisymmetric part of the $q$--th tensor power of the adjoint
representation of \g. This antisymmetric part is \g-reducible 
and $W_{2m_s-2}$
will have a fixed eigenvalue in each of its irreducible components 
(which  will change in general  when going from one
irreducible representation to another of the same or 
different ghost number). 
$s_{2m_s-2}$ and $\overline{s}_{2m_s-2}$  connect states belonging to the same 
\g-irreducible representation 
and with the same $W_{2m_s-2}$--eigenvalue 
(but of different ghost number) and such states will fall into one of the 
$\Sigma_{m_s}$ multiplets (singlets or doublets) discussed above. As the
generators $X_i$ commute with $*$, the \g-irreducible
representation  decomposition
pattern will be symmetrical under $q \rightarrow r-q$. 

The $l$ Casimir--Racah operators $\C^{(m_s)}$,
take fixed eigenvalues within each \g-irreducible component, 
which is uniquely labelled by them. 
The same irreducible
representation  may appear more than once, with equal or 
with different ghost numbers; the Casimirs will 
not distinguish among these
different copies of the same irreducible representation. 
As mentioned, $W_2 \equiv \Delta = -{1\over 2} \mathcal{C}^{(2)}$ 
(eqn. (\ref{2.32})).
An important question that naturally arises is whether 
$W_{2m_s-2}$ also reduces, for $s > 2$, to some higher order
Casimir-Racah operator or, more generally, to a sum of products of them.
Since $W_{2m_s-2}$ commutes with the $X$'s (realized in ghost space via 
(\ref{Xreal})) an equivalent question is
whether it belongs to the universal enveloping algebra 
$\mathcal{U}(\g)$ of \g.
The answer is negative and we address this
point in the next section, working out in full the 
$su(3)$ case.

\section{The case of $su(3)$}
\label{Csu3}
We opt here for mild departures from our previous
conventions: the generators will now be chosen 
hermitian so as to work with
real eigenvalues and the normalisation of all operators is such
that fractional eigenvalues are avoided. 
\subsection{Invariant tensors and operators}
\label{sec4.1}
The $su(3)$ algebra $ [ T_i, T_j ] = i \fijk T_k \, , \quad i=1, 
\ldots ,8$,
is determined by the non-zero structure constants $\fijk$ 
which are reproduced for convenience:
\begin{table}[H]
\caption{Non-zero structure constants for $su(3)$}
\label{tabfijk}
\begin{center}
\begin{tabular}{lll}
$f_{123}=1$ & $f_{147}=1/2$ & $f_{156}=-1/2$ \\
$f_{246}=1/2$ & $f_{257}=1/2$ & $f_{345}=1/2$ \\
$f_{367}=-1/2$ & $f_{458}=\sqrt{3}/2$ & $f_{678}=\sqrt{3}/2$
\end{tabular}
\end{center}
\end{table}
\noindent
These are also the coordinates of the
$su(3)$ three-cocycle.
The well known third order symmetric
tensor $d_{ijk}$ (table~\ref{tabdijk})
\begin{table}[H]
\caption{Non-zero components of the symmetric invariant $d_{ijk}$ for $su(3)$}
\label{tabdijk}
\begin{center}
\begin{tabular}{llll}
$d_{118}=1/\rtth$ & $d_{228}=1/\rtth$ &
               $d_{338}=1/\sqrt{3}$ & $d_{888}=-1/\rtth$ \\
$d_{448}=-1/(2\sqrt{3})$ & $d_{558}=-1/(2\sqrt{3})$ &
               $d_{668}=-1/(2\sqrt{3})$ & $d_{778}=-1/(2\rtth)$ \\
$d_{146}=1/2$ & $d_{157}=1/2$ & $d_{247}=-1/2$ & $d_{256}=1/2$ \\ 
$d_{344}=1/2$ & $d_{355}=1/2$ & $d_{366}=-1/2$ & $d_{377}=-1/2$
\end{tabular}  
\end{center}
\end{table}
\noindent
gives the third--order  Casimir--Racah operator.
{}From $d_{ijk}$ and (\ref{Omegacoord}) one finds the $su(3)$ 
five-cocycle coordinates \cite{Azc.Mac.Mou.Bue:97}
(table~\ref{tabOmega}).
\begin{table}[H]
\caption{Non-zero coordinates of the $su(3)$ five-cocycle}
\label{tabOmega}
\begin{center}
\begin{tabular}{lll}
$\Omega_{12345}=1/4$ & $\Omega_{12367}=1/4$ & 
                       $\Omega_{12458}=\sqrt{3}/12$ \\
$\Omega_{12678}=-\sqrt{3}/12$ & $\Omega_{13468}=-\sqrt{3}/12$ & 
                                $\Omega_{13578}=-\sqrt{3}/12$ \\
$\Omega_{23478}=\sqrt{3}/12$ & $\Omega_{23568}=-\sqrt{3}/12$ & 
                               $\Omega_{45678}=-\sqrt{3}/6$
\end{tabular}
\end{center}
\end{table}

The Casimirs $C_2$ and $C_3$,
\begin{equation}
C_2= T^i T_i 
\quad,\quad
C_3= d^{ijk} T_i T_j T_k
\quad ,
\label{C2C3def}
\end{equation}
are related to the operators (\ref{Cmdef}) simply by
\begin{equation}
C_2= -\C ^{(2)}  
\quad,\quad
C_3= -i\C ^{(3)} 
\quad .
\label{CcalCrel}
\end{equation}
The antisymmetric cocycles, on the other hand, give rise to the 
BRST and anti-BRST operators $s_2$, $\overline{s}_2$, $s_4$, 
$\overline{s}_4$ (see also (\ref{smjbdef}))

\begin{equation}
s_2 = - \frac{1}{2} \fijk c^i c^j \pi_k\quad,\quad
s_4= - \frac{1}{4!} {\Omega_{i_1 i_2 i_3 i_4}}^{\sigma} c^{i_1}
             c^{i_2} c^{i_3} c^{i_4} \pi_{\sigma} \quad ;
\label{s2s4def}
\end{equation}
\begin{equation}
{s_2}^2 = {s_4}^2 =0= {{\overline{s}_2}}^{\ 2} =
{{\overline{s}_4}}^{\ 2}  \quad,\quad
s_2 s_4 + s_4 s_2=0= \overline{s}_2 \overline{s}_4 + \overline{s}_4
\overline{s}_2 \ .
\label{s2s4cr}
\end{equation}
The corresponding Laplacians are
\begin{equation}
W_2=(s_2 + \overline{s}_2)^2 = s_2 \overline{s}_2 + \overline{s}_2 s_2
\quad,\quad
W_4=(s_4+\overline{s}_4)^2=s_4 \overline{s}_4 + \overline{s}_4 s_4 \,
\label{W2W4def}
\end{equation}
and satisfy, in addition to (\ref{2.27}),
\begin{equation}
[ W_2, \, (s_4,\overline{s}_4,W_4) \, ] =0  \quad, \quad
[W_4, \, s_4 ] =0= [ W_4, \, \overline{s}_4 ] \quad .
\label{W2W4s2s4}
\ee 
Notice that $W_4$ does not commute with $s_2$ or $\overline{s}_2$
(but, being invariant, it does commute with $W_2$).
As $W_2$ is proportional to $C_2$, we only refer to the latter in
the sequel.  Also, to avoid fractional eigenvalues, we define $W
\equiv 4! \, W_4$.

\subsection{Decomposition into irreducible representations}
\label{sec4.3}
In general, a monomial in the $c$'s of ghost number $q$ transforms in 
$\mathbf{8}^{\wedge q}$, 
the part of the $q$--th tensor power of the
$su(3)$ adjoint representation that is totally antisymmetric in the $q$ 
factors.
The reduction of the $\mathbf{8}^{\wedge q}$ into irreducible representations 
of $su(3)$ can be achieved by a variety of methods.
One way, which gives results useful 
in our analysis below, employs conventional tensor methods.
We first quote the results
\begin{equation}
\begin{array}{lclcl}
\mathbf{8}^{\wedge 0} & = & \mathbf{1} 
& = & \mathbf{8}^{\wedge 8} 
\\
\mathbf{8}^{\wedge 1} & = & \mathbf{8} 
& = & \mathbf{8}^{\wedge 7} 
\\
\mathbf{8}^{\wedge 2} & = & \mathbf{8} + \mathbf{10} + 
                          \overline{\mathbf{10}} 
 & = & \mathbf{8}^{\wedge 6}
\\
\mathbf{8}^{\wedge 3} & = & \mathbf{1} + \mathbf{8} + \mathbf{10} + 
                          \overline{\mathbf{10}} + \mathbf{27} 
 & = & \mathbf{8}^{\wedge 5}
\\
\mathbf{8}^{\wedge 4} & = & 2 \times \mathbf{8} +  
                         2 \times  \mathbf{27} 
\end{array}
\label{adjpow}
\end{equation}
noting the symmetry $\mathbf{8}^{\wedge q} = \mathbf{8}^{\wedge
(r-q)}$, and then describe the tensorial method of developing the results in 
fully explicit form.

We may refer to $su(3)$ irreducible representations either by dimension, or 
else in highest weight $\{\lambda_1,\lambda_2\}$ notation.
In the latter notation $\{1,0\}$ and $\{0,1\}$
denote the `quark' and `antiquark'
representations $\mathbf{3}$ and $\overline{\mathbf{3}}$ each of dimension 3,
and $\{\lambda_1,\lambda_2\}$
denotes the representation whose highest weight is
$\mathbf{w}(\lambda_1,\lambda_2)
= \lambda_1 \mathbf{w}(1,0) + \lambda_2 \mathbf{w}(0,1)$, where
$\mathbf{w}(1,0)$,  $\mathbf{w}(0,1)$ are the weights of 
$\mathbf{3}$, $\overline{\mathbf{3}}$ respectively.
The representation $\{\lambda_1,\lambda_2\}$ has dimension
\begin{equation}
d(\lambda_1,\lambda_2) = {1\over 2} (\lambda_1+1) (\lambda_2+1)
(\lambda_1+\lambda_2+2)
\end{equation}
and Casimir operators (\ref{C2C3def}) whose eigenvalues are
\cite{Bie:63}
\begin{equation}
C_2(\lambda_1,\lambda_2)= \frac{1}{3} (\lambda_1^2 +\lambda_1\lambda_2 
+\lambda_2^2) +\lambda_1 +\lambda_2 \, \, .
\label{C2pq}
\end{equation}
\begin{equation}
C_3(\lambda_1,\lambda_2)= {1\over 18} (\lambda_1-\lambda_2)
(\lambda_1+2\lambda_2+3)(2\lambda_1+\lambda_2+3) = - C_3(\lambda_2,\lambda_1)
\, \, .
\label{C3pq}
\end{equation}
Since $C_3(\lambda_1,\lambda_2)=-C_3(\lambda_2,\lambda_1)$, $C_3$ vanishes for
all self-conjugate ($\lambda_1=\lambda_2$) irreducible representations.
The results for the representations $\rho$ that occur in (\ref{adjpow}) are 
given in the table
\begin{equation}
\begin{array}{c@{\quad\quad}c@{\quad\quad\quad}c@{\quad\quad\quad}cc}
\dim \rho   		& & \{\lambda_1,\lambda_2\} & (C_2, \, C_3)
\\[0.4cm]
\mathbf{1} 		& & \{0,\,0\} & (0, \, 0)
\\[0.3cm]
\mathbf{8} 		& & \{1,\,1\} & (3, \, 0)
\\[0.3cm]
\mathbf{10} 		& & \{3,\,0\} & (6, \, 9)
\\[0.3cm]
\overline{\mathbf{10}}  & & \{0,\,3\} & (6, \, -9)
\\[0.3cm]
\mathbf{27} 	& & \{2,\,2\} & (8, \, 0) &\quad .
\end{array}
\label{C2C3lab}
\end{equation}

Turning to the tensor analysis of tensors spanned, for $0\le q\le 8$, by the 
monomials $c_{i_1}\cdots c_{i_q}$, we start with the case $q=1$, where $c_i$ 
describes the basis of the $su(3)$ adjoint representation, \ie, an octet.
In the case $q=2$, 
\begin{equation}
d_{i} = f_{ijk} c_j c_k
\label{octet2}
\end{equation}
describes an independent octet, the only one 
available since $d_{ijk} c_j c_k\equiv 0$.
The remaining tensor, irreducible over the field $\mathbb{R}$, is
\begin{equation}
c_i c_j - {1\over 3} f_{i j k} d_k = (\mathbf{20}_2)_{ij}\quad,
\label{dec2}
\end{equation}
for which $f_{ijk} (\mathbf{20}_2)_{ij} =0$ by construction.
The notation implies it has 20 components, agreeing with the simple count
${8\choose 2} - 8$.
To reduce it into separate $\mathbf{10}$ and
$\overline{\mathbf{10}}$ pieces can 
be done only over the field of complex numbers, but this is not needed here
\cite{Sud:69}.
We may also write
\begin{equation}
\begin{array}{rl}
c_i c_j = & \displaystyle
        (c_i c_j - {1\over 3} f_{i j k} d_k) + {1\over 3} f_{i j k} d_k
\\[0.3cm]
= & \displaystyle
(P_{20})_{ij,pq} c_p c_q + (P_{8})_{ij,pq} c_p c_q
\end{array}
\end{equation}
where the projectors are given by
\begin{equation}
\begin{array}{l}
\displaystyle
(P_{20})_{ij,pq} = {1\over 2} (\delta_{ip}\delta_{jq} - \delta_{iq} 
\delta_{jp}) - {1\over 3} f_{i j l}  f_{l p q}
\\[0.3cm]
\displaystyle
(P_8)_{ij,pq}= {1\over 3} f_{i j l}  f_{l p q}
\quad.
\end{array}
\end{equation}
The projection properties and orthogonality can be checked using well known 
properties of $su(3)$ $f$-tensors etc.
\cite{Azc.Mac.Mou.Bue:97,Mac.Sud.Wei:68}.
Also we have trivially
\begin{equation}
P_{20} + P_8 = U
\quad,\quad
U_{ij,pq} = {1\over 2} (\delta_{ip}\delta_{jq} - \delta_{iq} \delta_{jp})
\end{equation}
where $U$ is the relevant form of the unit operator in the ghost number
$q=2$ space spanned
by the antisymmetric tensors $c_i  c_j$.
Direct calculations on the explicit form for $d_k$ given by (\ref{octet2}) 
and for
$(\mathbf{20}_2)_{ij}$ by (\ref{dec2}) show 
that these have the $C_2$ eigenvalues $3$ and $6$ in
(\ref{C2C3lab}).

The space spanned at $q=3$ by the tensor components $c_i c_j c_k$ gives rise 
easily to the singlet (0,0)
\begin{equation}
Y = f_{ijk} c_i c_j c_k = c_i d_j
\label{3scalar}
\end{equation}
and the $su(3)$ octet
\begin{equation}
e_i = d_{ijk} c_j d_k
\quad.
\label{3octet}
\end{equation}
This is the only $q=3$ octet, since
\begin{equation}
\xi_i = f_{ijk} c_j d_k =0
\end{equation}
follows from the definition (\ref{octet2}) of $d_k$ and the Jacobi identity 
for the $f$-tensor.
We note however that $\xi_i=0$ is a set of eight non-empty 
verifiable identities amongst various trilinears $ c_i c_j c_k$.
To build other irreducible tensors, it is natural to look at the tensors
\begin{equation}
c_i d_j - c_j d_i
\label{3rdskew}
\end{equation}
\begin{equation}
c_i d_j + c_j d_i
\label{3rdsym}
\end{equation}
with a priori 28 and 36 components.
The former (\ref{3rdskew}) is irreducible and defines $(\mathbf{20}_3)_{ij}$
as it stands, because $\xi_i=0$ yield eight identities automatically satisfied 
by its components.
It is also not hard to check that the $C_2$ eigenvalue is 6.
The latter (\ref{3rdsym}) is not irreducible, but by extracting the scalar 
(\ref{3scalar}) and the octet (\ref{3octet}), we find the irreducible tensor of 
27= 36--1--8 components
\begin{equation}
(\mathbf{27}_3)_{ij} = c_i d_j + c_j d_i - {1\over 4} \delta_{ij} Y
- {6\over 5} d_{ijk} e_k\quad.
\label{327}
\end{equation}

It is easy to see that contracting with $\delta_{ij}$ and $d_{ijk}$ gives zero 
as irreducibility requires.
It is hard, needing good selection of $su(3)$ such $f$- and $d$-tensor 
identities as found in \cite{Azc.Mac.Mou.Bue:97},
to prove that $C_2$ indeed has eigenvalue 8 for
$(\mathbf{27}_3)_{ij}$.
We could turn results (\ref{3scalar}), (\ref{3octet}), (\ref{3rdskew}) and 
(\ref{327}) into the form
\begin{equation}
c_i c_j c_k = \sum_R P^R_{ijk,pqr} c_p c_q c_r
\end{equation}
involving a complete set of orthogonal projectors for $R=1,\ 8,\ 20$ and 27.

Since the case at $q=4$ involves repetitions,
it is best at this point to review the situation regarding octets.
At $q=1,\ 2,\ 3$, we have
\begin{equation}
c_i\quad,\quad d_i\quad,\quad e_i
\end{equation}
and no others.
At $q=4$, we find
\begin{equation}
f_i = d_{i j k} d_j d_k = \Omega_{i\, i_1 i_2 i_3 i_4} c_{i_1} c_{i_2} c_{i_3} 
c_{i_4} 
\quad,
\label{4octet}
\end{equation}
but $f_{ijk} d_j d_k\equiv 0$.
A second octet that can be checked easily to be linearly 
independent of $f_i$ is
$Y c_i$.
We may build other $q=4$ octets, but these will not give anything
new, since \eg \ we can prove the results
\begin{equation}
d_{ijk} c_j e_k = -{2 \over 3} c_i Y \quad , \quad \quad
f_{ijk} c_j e_k = f_i \quad .
\label{dcefce}
\end{equation}
It is thus now obvious that the complete family of octets 
can be presented as
\begin{equation}
\begin{array}{cccccccc}
q = & 1   & 2   & 3   & 4     & 5     & 6     & 7 
\\[0.3cm]
    & c_i & d_i & e_i & f_i
\\[0.3cm]
    &     &     &     & Y c_i & Y d_i & Y e_i & Y f_i \quad.
\end{array}
\label{4.22}
\end{equation}
As a check, we find that $Y f_i$ (\eg) is as expected,
\begin{equation}
Y f_1 \sim c_2 c_3 c_4 c_5 c_6 c_7 c_8 \sim * c_1 \quad.
\end{equation}

An alternative but equivalent treatment would employ 
certain duals of $f_i,\ 
e_i,\ d_i,\ c_i$ in place of $Y c_i,\ Y d_i,\ Y e_i, Y f_i$ in 
(\ref{4.22}).
Of course, for the last case, we have just proved the easy bit of the 
equivalence.
In fact, the use of duals in explicit work is much less 
convenient than the 
choice used in (\ref{4.22}).
To indicate this, and to do something instructive in its own right, we 
make explicit the dual relation of $f_i$ and $Y c_i$
(see also \cite[section~8]{Azc.Mac.Mou.Bue:97}).

To make contact with a dual to the octet $f_i$ of (\ref{4octet}) replace
$c_{i_1} c_{i_2} c_{i_3} c_{i_4}$ there by
$\epsilon_{i_1 i_2 i_3 i_4 j_1 j_2 j_3 j_4} 
c_{j_1} c_{j_2} c_{j_3} c_{j_4}$
to reach
\begin{equation}
p_i = \Omega_{i\, i_1 i_2 i_3 i_4}
\epsilon_{i_1 i_2 i_3 i_4 j_1 j_2 j_3 j_4} c_{j_1} 
c_{j_2} c_{j_3} c_{j_4}
\quad,
\end{equation}
which clearly belongs to $\mathbf{8}^{\wedge 4}$.
To relate $p_i$ to $Y c_i$, we need the identity
\begin{equation}
{1\over 4!}
\Omega_{i \, i_1 i_2 i_3 i_4} \epsilon_{i_1 i_2 i_3 i_4 j_1 j_2 j_3 j_4}
= {2\over \sqrt{3}} \delta_{i [j_1} f_{j_2 j_3 j_4]}
\label{4.24}
\end{equation}
in which the divisor $4!$ on the left is actually matched by 
one implicit in 
our definition of square antisymmetrisation brackets on the right.
Identity (\ref{4.24}) allows us to prove
\begin{equation}
p_i = {4!\cdot 2\over \sqrt{3}} c_i Y = 
 - {4!\cdot 2\over \sqrt{3}} Y c_i
\quad,
\end{equation}
as expected.

The contraction $i= j_1$ of (\ref{4.24}) gives
\begin{equation}
{1\over 4!}
\Omega_{i_1 i_2 i_3 i_4 i_5} \epsilon_{i_1 i_2 i_3 i_4 i_5 j_1 j_2 j_3}
= {5\over 2\sqrt{3}} f_{j_1 j_2 j_3}
\label{4.26}
\end{equation}
which is an evident and easily checked analogue of the result
\begin{equation}
{1\over 3!}
f_{j_1 j_2 j_3} \epsilon_{j_1 j_2 j_3 i_1 i_2 i_3 i_4 i_5}
= - {2\sqrt{3}} \Omega_{i_1 i_2 i_3 i_4 i_5}
\label{4.27}
\end{equation}
given in \cite[eqn.~(8.14)]{Azc.Mac.Mou.Bue:97}.
The latter is a contraction of the more useful identity
\begin{equation}
{1\over 3!}
f_{i j_1 j_2} \epsilon_{j_1 j_2 i_1 i_2 i_3 i_4 i_5 i_6}
= - {4\sqrt{3}} \delta_{i [i_1} \Omega_{i_2 i_3 i_4 i_5 i_6]}
\quad.
\label{4.28}
\end{equation}
This may be used, as we used (\ref{4.24}), to reach, \eg, eventually 
the dual relationship of $d_i$ to $Y e_i$. 

Whilst the above tensorial analysis provides an explicit construction from 
first principles of all the entries of (\ref{adjpow}) the use of $s_2$ and 
$s_4$ expedites explicit work.
For example, since $[s_2, X_i] =0= [s_4, X_i]$, $s_2$ and $s_4$
also commute with $C_2$ and $C_3$. Thus, $s_2$ (\eg) 
either raises the ghost number of a tensor by one, 
leaving its $su(3)$ nature unaltered or else annihilates it.
Thus $s_2 c_i \sim d_i$, $s_2 d_i = 0$.
Similarly, $s_4 c_i \sim f_i$ and $\overline{s}_2 s_4 f_i \sim e_i$.
Likewise,
\begin{equation}
s_2(\mathbf{20}_2)_{ij} = s_2(c_i c_j - {1\over 3} f_{ijk} d_k)
\sim
d_i c_j - c_i d_j = -  (\mathbf{20}_3)_{ij}
\quad,
\end{equation}
since $s_2 d_i = 0$, which confirms what has been seen to hold above.

Further we might expect $s_2 (\mathbf{27}_3)_{ij}$ to yield one of the required
$(\mathbf{27}_4)_{ij}$.
Indeed $s_2 d_i = 0$, $s_2 Y=0$ and $s_2 e_k = f_k$ allow us to write
\begin{equation}
s_2 (\mathbf{27}_3)_{ij} = d_i d_j - {3\over 5} d_{ijk} f_k \equiv
(\mathbf{27}_4)_{ij}
\label{4.30}
\quad.
\end{equation}
A second 27-tensor in $\mathbf{8}^{\wedge 4}$ that is linearly independent of 
$(\mathbf{27}_4)_{ij}$ of (\ref{4.30}) is suggested immediately by duality 
arguments.
One replaces $c_{i_1} c_{i_2} c_{i_3} c_{i_4}$ in $d_i d_j =
f_{i i_1 i_2} f_{j i_3 i_4} c_{i_1} c_{i_2} c_{i_3} c_{i_4}$, etc. by
$\epsilon_{i_1 i_2 i_3 i_4 j_1 j_2 j_3 j_4} c_{j_1} c_{j_2} c_{j_3} c_{j_4}$.
We thereby reach a tensor $(\mathbf{27}_4')_{ij}$ which is 
plainly linearly independent of $(\mathbf{27}_4)_{ij}$.
It turns out to be proportional to 
\begin{equation}
(\mathbf{27}_4')_{ij} = c_i e_j + c_j e_i - {4\over 5} d_{ijk} Y c_k
\quad,
\end{equation}
which can be seen to satisfy $d_{ijl} (27_{4}^{\prime})_{ij}=0$,
using (\ref{dcefce}), as well as $(27_{4}^{\prime})_{ii}=0$, so
that it is irreducible, with 27 components. Further,
$(27_{5})_{ij}$ can now be written down explicitly by action of
$s_2$ on $(27_{4}^{\prime})_{ij}$.  
No systematic work on projectors for 
$q=4$ has been done.

\subsection{The Laplacian $W$}
{}From the analysis of section~\ref{sec4.3} 
of the $su(3)$ representations contained 
in $\mathbf{8}^{\wedge q}$, $0\le q\le 8$, where $q$ is the ghost 
number, it 
can be seen that the states $\varphi$
of the system are labelled by the eigenvalues of 
the ghost number operator $Q=c^i \pi_i$,
$Q\varphi=q\varphi$, and of the $su(3)$ Casimirs $C_2$ and $C_3$
that label states within each $su(3)$ representation.
Since $W$ commutes with $Q,\ C_2,\ C_3$, we expect it to have 
well defined 
eigenvalues on all the states of the system and we might 
further expect it to 
be defined as a specific function of $Q,\ C_2,\ C_3$.

Some progress can be made analytically to compute $W$-eigenvalues.
For example, for $c_i$ and $d_i$ given by (\ref{octet2})
which describe $q=1,\ q=2$ octets, we may compute directly from 
(\ref{W2W4def}) the results
\begin{equation}
W c_i = 5 c_i\quad,\quad W d_i =0
\quad.
\label{4.3.0}
\end{equation}
These calculations, the latter already non-trivial, depend, amongst other 
things, on the identities
\begin{equation}
\begin{array}{c}
\Omega_{i_1 i_2 i_3 i_4 p} \Omega_{i_1 i_2 i_3 i_4 q} = 5 \delta _{pq}
\quad,
\\[0.3cm]
\displaystyle
\Omega_{i_1 i_2 i_3 a b} \Omega_{i_1 i_2 i_3 p q} = 
{1\over 2} (\delta_{ap} \delta_{bq} - \delta_{aq} \delta_{bp} 
+ f_{ab i} f_{pq i})
\quad,
\end{array}
\label{4.3.1}
\end{equation}
of which only the first follows from the definition of $\Omega$ easily.
Note also that since $W$ distinguishes between different octets,
eqn. (\ref{4.3.0}), it cannot be a pure 
function of the Casimirs: it depends also on $Q$, which does not 
belong to the $\mathcal{U}(su(3))$ enveloping algebra.

The results of
section~\ref{sec4.1} also allow the minimal polynomials for $C_2$ and 
$C_3$ to be deduced.
These are
\begin{equation}
C_2(C_2-3)(C_2-6)(C_2-8) =  0 \quad ,
\label{C2minpol}
\end{equation}
\begin{equation}
C_3(C_3+9)(C_3-9) =0 \quad,
\label{C3minpol}
\end{equation}
and the orthogonal projectors on the various eigenspaces for $C_2$ and $C_3$ are
\begin{equation}
\begin{array}{@{}l@{}l@{}l}
\displaystyle
P^{(2)}_0= - \frac{1}{144} 
(C_2  -  3)(C_2  -  6)(C_2  -  8)
\ \ &
\displaystyle
P^{(3)}_0= - \frac{1}{81} (C_3  +  9)(C_3  -  9) \ \ & 
\\[0.3cm]
\displaystyle
P^{(2)}_3=\frac{1}{45} C_2 (C_2-6)(C_2-8) & 
\displaystyle
P^{(3)}_{-9}= \frac{1}{162} C_3(C_3-9) & 
\\[0.3cm]
\displaystyle
P^{(2)}_6=-\frac{1}{36} C_2 (C_2-3)(C_2-8) &
\displaystyle
P^{(3)}_9=\frac{1}{162} C_3 (C_3 +9) \;. &
\\[0.3cm]
\displaystyle
P^{(2)}_8=\frac{1}{80} C_2 (C_2-3)(C_2-6) \;;&
\end{array}
\label{projopdef}
\end{equation}

Further progress by analytic methods soon becomes difficult and we have made 
use of FORM \cite{Ver:92}.
This enables us firstly to compute all $W$-eigenvalues, discussed below, and 
to find the following identities
\begin{equation}
\begin{array}{@\displaystyle r @{} c @{} @\displaystyle l}
C_2 C_3 & = & 6 C_3 \\[0.3cm]
{C_3}^2 & = & \displaystyle -\frac{9}{4} C_2 (C_2-3)(C_2-8) \\[0.3cm]
C_2 W & = & 3 W \displaystyle - \frac{1}{2} C_2 (C_2-3)(C_2-8) \\[0.3cm]
W^2 & = & \displaystyle 5 W + \frac{2}{27} {C_3}^2 \, \,  .
\end{array}
\label{FORMres}
\end{equation}
These results allow the recovery of (\ref{C2minpol}), 
(\ref{C3minpol}) as a mild 
check on our procedures, and the deduction of the minimal polynomial of $W$
\begin{equation}
W(W-5)(W-6)=0 \quad , 
\label{Wminpol}
\end{equation}
which comprises, as it should, all the eigenvalues of $W$ found in practice.
Also the orthogonal projectors onto the eigenspaces of $W$ are
\begin{equation}
P^{(W)}_0=\frac{1}{30} (W  -  5)(W  -  6) \quad,\quad
P^{(W)}_5=-\frac{1}{5} W(W-6) \quad,\quad
P^{(W)}_6= \frac{1}{6} W (W-5)\quad.
\label{projopdef1}
\end{equation}
Various useful inferences can be made regarding eigenspaces.
For example, alongside the previous result
$\text{ker}\,C_2\subseteq \text{ker}\,W$, we have
$\text{ker}\,W\subseteq \text{ker}\,C_3$.
Also,
\begin{eqnarray}
P^{(3)}_9 + P^{(3)}_{-9} & = & P^{(2)}_6 = P^{(W)}_6 \nonumber \\
P^{(2)}_0 P^{(W)}_0 & = & P^{(2)}_0 \nonumber \\
P^{(2)}_8 P^{(W)}_0 & = & P^{(2)}_8 \nonumber \\
P^{(W)}_0 + P^{(W)}_5 & = & P^{(3)}_0 \quad .
\label{projoprel}
\end{eqnarray}

So far no explicit expression for $W$ in terms of $Q,\ C_2,\ C_3$ is at hand.
The major complication in the pattern of the $W$-eigenvalues of the $su(3)$ 
representations in $\mathbf{8}^{\wedge q}$ concerns the octets.
For these, the ghost number $q=1,4,4,7$ octets have eigenvalue $W=5$ and the
$q=2,3,5,6$ 
octets $W=0$.
This suggests the use of Lagrangian interpolation to define a function
\begin{equation}
f(q)={1\over 360}[ (q-4)^2 -10(q-7)(q-1)](q-2)(q-3)(q-5)(q-6) \quad,
\label{4.47}
\end{equation}
which equals 1 at $q=1,4,4,7$ and 0 at $q=2,3,5,6$, so that for these values,
$f(q)^2=f(q)$.
It is then immediate to see that the formula
\begin{equation}
W={1\over 9} C_2 (C_2 -6)(C_2-8) f(Q) - {1\over 6} C_2 (C_2-3) (C_2-8) =
5 P^{(2)}_3 f(Q) + 6 P^{(2)}_6 \quad,
\label{4.48}
\end{equation}
correctly predicts the $W$-eigenvalues of all states.
The last two equations of (\ref{FORMres}) also follow directly, using 
projector properties,
$f(Q)^2=f(Q)$ and the second equation of (\ref{FORMres}).
We should stress here that $f(Q)^2=f(Q)$ holds only for $q\in \{1,2,\dots,7\}$
whereas the allowed range of values of $q$ is $\{0,1,\dots,8\}$.
But this does not matter for (\ref{4.48}), because although $f(Q)$ is finite 
$(=-27)$ at $q=0$ and $q=8$, $C_2=0$ for the $q=0$ and $q=8$ states.
Finally, FORM confirms that $W$ defined by (\ref{4.48}), and 
$W=4! W_4$ given by (\ref{W2W4def}) are equal as operators.

A final remark about the form of (\ref{4.47}), (\ref{4.48}) is in
order here. Observing that $f(Q) = f(8-Q)$, one is led to write
(\ref{4.47}) in terms of $u:= Q(8-Q)$, finding 
\begin{equation}
f(Q)=F(u):= \frac{1}{40} (u-6)(u-12)(u-15) 
\label{Fu}
\end{equation}
(a form that can also be directly derived by Lagrangian interpolation).
A different approach is to start from the minimal polynomial for
$u$
\begin{equation}
u(u-7)(u-12)(u-15)(u-16)=0
\label{uminpoly}
\end{equation}
and write down directly a function $\tilde{F}(u)$, 
with $\tilde{F}(u)=0$ at
$u=0,12,15$ and $\tilde{F}(u)=1$ at $u=7,16$ 
\begin{equation}
\tilde{F}(u)=P^{(u)}_7 + P^{(u)}_{16} \, ,
\label{Fpu}
\end{equation}
the projectors $P^{(u)}_\lambda$ being defined in the standard way
from (\ref{uminpoly}). Using this $\tilde{F}(u)$ in place 
of $F(u)=f(Q)$ in
(\ref{4.48}) also gives correctly $W$ - the difference $\tilde{F}-F$
is annihilated by $P^{(2)}_3$ \footnote{ We note incidentally that 
the operators $M_{ij}:= c_i \pi_j - c_j \pi_i$, 
$i <j$, generate the algebra $spin(8)$, the quadratic Casimir of
which is proportional to $u$ (since $M_{ij} M_{ij}=-2u$), \ie 
\  (\ref{4.48})
gives $W$ in terms of the quadratic Casimirs of $su(3)$ and
$spin(8)$.}.

The results of the previous analysis may be summarised in
the diagram of figure~\ref{piechart} representing the 
spectra of $C_2$, $C_3$ and $W$.
\begin{figure}
\centerline{\includegraphics{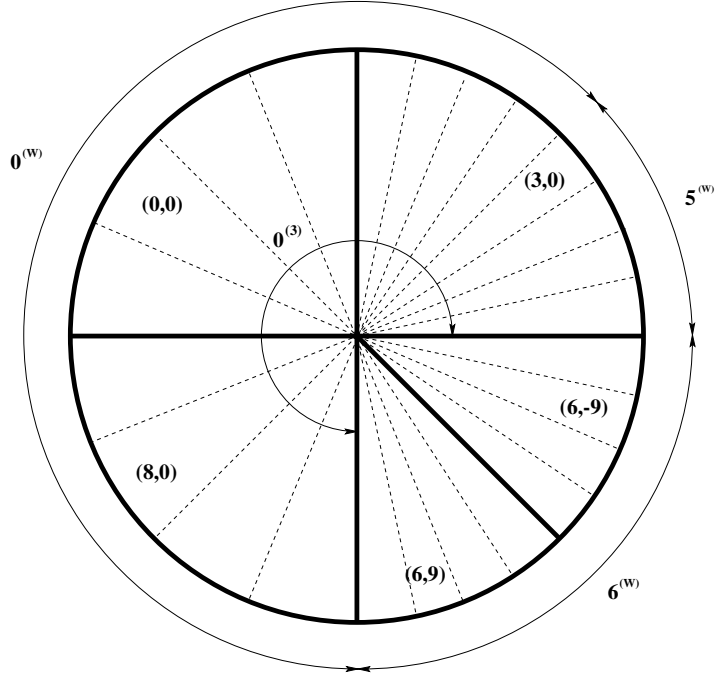}}
\caption{The spectrum of the Casimirs ($C_2$, $C_3$) and of 
$W_4 \propto W$}
\label{piechart}
\end{figure}
\noindent
The solid circular disc represents all the $2^8=256$ states
available in $\bigoplus_{i=0}^8 \mathbf{8}^{\wedge i}$. The four 
quadrants represent the four eigenspaces $(0,3,6,8)$ of
$C_2$ while the numbers in parentheses are the eigenvalues of
$C_2$, $C_3$ valid in each disc segment (bordered by solid black
lines). Dashed lines within each disc segment separate multiple
copies of the same irreducible representation, corresponding in
general to states of different ghost number. The arcs outside the disc
specify the three (0,5,6) $W$--eigenspaces. 
We summarise its key features:
\begin{itemize}
\item
The eigenspace $\text{Im}\,P^{(2)}_6$, equal
to $\text{Im}\,P_6^{(W)}$, is
split into two parts with the same number of states, labelled by 
the $C_3$ eigenvalues 9 and
-9.
With the help of (\ref{C2C3lab}) we recognise these, respectively, as the
$\mathbf{10}$ and $\overline{\mathbf{10}}$
$su(3)$ representations,
each of which appears four times,
with ghost numbers 2, 3, 5 and 6.
\item
The $(0, \, 0)$--subspace (\emph{i.e.} $\ker C_2 = \K_2$), 
contains
four invariant states (the $su(3)$--singlets $\mathbf{1}$), 
with ghost numbers 0, 3, 5 and 8. All of them are $W$--harmonic as
well, \ie, $\Sigma_4$ singlets.
\item
$(8, \, 0)$ ($=\mathbf{27}$) appears four times, with ghost numbers
3, 5 and 4 (twice) and is also $W$--harmonic.
\item $(3, \, 0)$ ($=\mathbf{8}$) appears eight
times, with ghost numbers 1 through 7 (twice for 4).
\end{itemize}
The diagram shows that 
$\K_2 + \text{Im}\, P^{(2)}_8 \subset \K_4 \subset \ker
C_3$. $\K_4$ contains half of the copies of the adjoint 
representation, the rest belonging to the $W$--eigenvalue 5.

To look now at the representations of $\Sigma_2$ and $\Sigma_4$
in (\ref{2.27}) and (\ref{ssbWalg}) it is convenient to
depict the $su(3)$ representations as in the diagram of figure~\ref{ladder}
and to analyse there the role played by
$s_2$, $s_4$ and their adjoints $\overline{s}_2$, $\overline{s}_4$ 
in interconnecting them.
\begin{figure}
\centerline{\scalebox{1.3}{\includegraphics{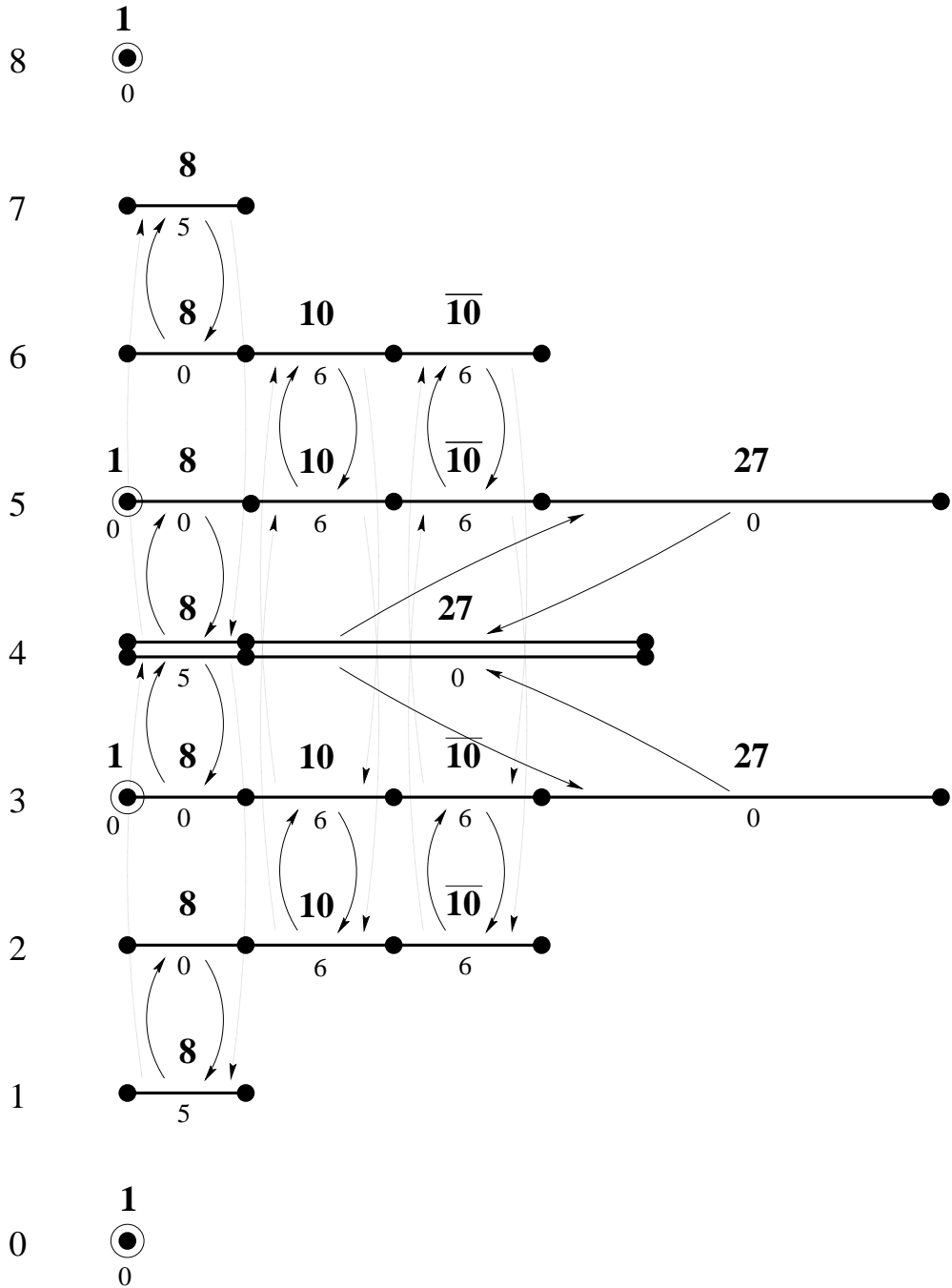}}}
\caption{Decomposition of
$\bigoplus_{i=0}^8 \mathbf{8}^{\wedge i}$ into irreps 
of $su(3)$ and the action 
of $s_2$ ($\overline{s}_2$) and $s_4$ ($\overline{s}_4$) on 
them (notice that $[s_2\, (\overline{s}_2), W]\neq 0$).
The eigenvalue of $W$ appears under each irrep.}
\label{ladder}
\end{figure}
Each \emph{straight line} segment in this diagram represents an 
irreducible representation.
All segments in the same line correspond to the same ghost number,
while singlets are represented by circles. The
arrows between segments depict the action of the $s$'s (solid (black)
lines for $s_2$, $\overline{s}_2$, dotted (grey)
lines for $s_4$, $\overline{s}_4$); the
number below each segment is the $W$--eigenvalue of the irreducible
representation. In
the following we denote \emph{e.g.} by $\mathbf{10}_3$ the irreducible
representation $\mathbf{10}$
with ghost number 3 while a further superscript $u$ ($l$)  
(for \emph{upper} (\emph{lower})) distinguishes
between the two ghost number 4 representations $\mathbf{8}_4$'s and 
$\mathbf{27}_4$'s.
We point out the following
\begin{itemize}
\item 
Referring to multiplets of the superalgebra (\ref{ssbWalg}),
quartets--turned--into--pairs--of--doublets, according to the
remark of section~\ref{RhoHd}, appear three times. For $\Sigma_2$, 
the quartet $\{ \mathbf{8}_4^u, \, \mathbf{8}_5, \, \mathbf{8}_3, \,
\mathbf{8}_4^l \}$ actually consists of the pair of doublets 
$\{ \mathbf{8}_5, \,
\mathbf{8}_4^u \}$, $\{ \mathbf{8}_4^l, \, \mathbf{8}_3 \}$
---a similar pattern
is exhibited by the $\mathbf{27}$'s in the same orders as well as
by the \hbox{$\Sigma_4$--quartet} $\{ \mathbf{8}_4^u, \, \mathbf{8}_7, \,
\mathbf{8}_1, \, \mathbf{8}_4^l \}$. 
The degeneracy seen in the $q=4$ line is then resolved by noting that $s_2$ 
annihilates one octet and $\overline{s}_2$ the other (and similarly for the 
$\mathbf{27}$).

\item
Besides the above `split quartets', we also have the $\Sigma_2$--doublets 
$\{\mathbf{8}_2\,,\,\mathbf{8}_1\}$, $\{\mathbf{10}_3\,,\,\mathbf{10}_2\}$,
$\{\overline{\mathbf{10}}_3\,,\,\overline{\mathbf{10}}_2\}$, the 
$\Sigma_4$--doublets
$\{\mathbf{10}_5\,,\,\mathbf{10}_2\}$,
$\{\overline{\mathbf{10}}_5\,,\,\overline{\mathbf{10}}_2\}$,
and their $*$--images
$\{\mathbf{8}_7\,,\,\mathbf{8}_6\}$, $\{\mathbf{10}_6\,,\,\mathbf{10}_5\}$,
$\{\overline{\mathbf{10}}_6\,,\,\overline{\mathbf{10}}_5\}$  
and 
$\{\mathbf{10}_6\,,\allowbreak\mathbf{10}_3\}$,
$\{\overline{\mathbf{10}}_6\,,\,\overline{\mathbf{10}}_3\}$ 
respectively.
 Notice that $W$ changes eigenvalue within all $\Sigma_2$
doublets involving $\mathbf{8}$'s, reflecting its failure to
commute with $s_2$.

\item 
The $su(3)$ (and hence $\Sigma_2,\Sigma_4$)
singlet $\mathbf{1}_0$ is simply the constant monomial 1,
while $\mathbf{1}_3$ is the three-cocycle $f_{ijk} c^i c^j c^k=Y$. The
other two singlets are the `top form' $c_{1} \ldots c_{8}$
and the five-cocycle $\Omega_{i_1\dots i_5} c^{i_1}\dots c^{i_5}$,
$*$--images of the first two respectively.
\item $\mathbf{8}_1$
consists of the 8 $c^k$'s. This is `lifted' by $s_2$ to give 
 $\mathbf{8}_2 \sim \{ \fijk c^i c^j \}$ and by $s_4$, giving
$\mathbf{8}_4^l \sim \{ {\Omega_{i_1 i_2 i_3 i_4}}^k c^{i_1}
c^{i_2} c^{i_3} c^{i_4} \}$. $\mathbf{8}_3$ is the image of
$\mathbf{8}_1$ under $\overline{s}_2 s_4$, \emph{i.e.} 
$\mathbf{8}_3 \sim 
\{ f_{i_1 a b} {\Omega_{i_2 i_3}}^{abk} c^{i_1} c^{i_2} c^{i_3}
\}$. The $q \rightarrow r-q$ symmetry accounts for the rest of the
$\mathbf{8}$'s. Notice that $\mathbf{8}_2$, $\mathbf{8}_3$ cannot 
be lifted by $s_4$ since they are $W$--harmonic. 
\end{itemize}
\section{Concluding remarks}
\label{Conclusion}

We have introduced and studied in this paper the supersymmetry algebra 
generated by the higher order BRST operators.
The central term in the algebra is given, in the standard (lowest
order) case,  by the 
(quadratic) Casimir. 
As shown explicitly by the expression of $W_4$ for the 
algebra $\g=su(3)$, the 
higher order Laplacians may
involve the ghost number operator, which, unlike the 
Casimir-Racah operators, lies outside the enveloping algebra 
$\mathcal{U}(\g)$.
Thus, the fact that $\Delta\in\mathcal{U}(\g)$ in the 
standard case is rather exceptional.

We wish to conclude with a purely mathematical remark.
Using the correspondence $c^i\leftrightarrow$LI forms on the group 
manifold, the standard BRST operator $s_2$ may be identified with 
the exterior derivative $d$ acting on forms.
The basic properties of $d$, $d:\wedge^q\to\wedge^{q+1},\ d^2=0$ (and 
of the codifferential $\delta$) may be extended by introducing 
generalised operators $\tilde d$ in two different ways.
One is by replacing the exterior differential by a higher order nilpotent 
endomorphism $\tilde d'$ satisfying $(\tilde d')^k=0$, to study 
the associated generalised homology, etc. \cite{Dub:96,Ker:97}.
This approach is reminiscent of the one used to generalise ordinary 
supersymmetry to fractional supersymmetry (for a review with earlier 
references see \cite{Dun.Mac.Azc.Bue:97}). The second one 
replaces $d$ by a $p$-th order differential, $\tilde d$, $p$ 
odd, satisfying $\tilde d_p:\wedge^q\to\wedge^{q+p},
\ \tilde d_p^2=0$, and it is this
second point of view which
corresponds to the analysis presented in this paper.
In fact, the higher BRST operator $s_{2m-3}$
may be considered as an explicit 
construction of this differential
for $p$=$(2m-3)$, which acts \emph{on LI} forms on 
the group manifold by translating (\ref{genmc}) using the 
above correspondence.
$\tilde d_p$ is an odd operator satisfying
\begin{equation}
\tilde d_p (\alpha\wedge\beta)= (\tilde d_p \alpha) \wedge \beta 
+ (-1)^n \alpha\wedge (\tilde d_p \beta )
\label{holeibniz}
\end{equation}
where $n$ is the order of the LI form $\alpha$.
We recall that, using the (standard) product 
between manifolds and forms (given by  
$\langle \mathcal{M},  \alpha\rangle = \int_\mathcal{M} \alpha$)
one can define the adjoint $\partial$ of the exterior derivative $d$.
Acting on manifolds, $\partial$ reduces their dimension by one, is 
nilpotent, and admits the interpretation as a boundary
operator. Using an analogous procedure, one might think of defining
the adjoint $\tilde{\partial}_p$ of $\tilde{d}_p$ as an operator. 
Acting on manifolds it would reduce  their dimension by $p$, being 
also nilpotent, and the question would arise whether it, too, 
admits a simple topological interpretation. One might also ask 
further, whether an analogue of Stokes' theorem could be 
formulated along these lines or whether 
the spectrum of the higher order Laplacians studied here 
provides topological information about the underlying manifold.
We do not know whether these mathematical constructions
involving $\tilde{d}_p$ can be carried through in general.

To conclude we would like to stress that the 
cohomological properties used in this paper are
also relevant in other related fields, although it may
not be directly apparent. They determine and classify, 
for instance, the local conserved charges in principal 
chiral models (see \cite{Eva.Has.Mac.Mou:99} and
references therein), and are also important in 
$W$-algebras (see, for instance, \cite{Bai.Bou.Sur.Sch:88, 
Ber.Bil.Ste:91, Lu.Po.Wa.Xu:94} and~\cite{Bou.Car.Pil:96}), 
where BRST-type techniques, and hence Lie algebra cohomology, 
are relevant.

\subsection*{Acknowledgements}
This research is partially supported by a research grant PB96-0756
from the DGICYT,  Spain. C.C. wishes to thank the Spanish
Ministry of Education and Culture for a post-doctoral fellowship  
and J.C.P.B.  wishes to thank the Spanish MEC and CSIC for an FPI grant.
A.J.M wishes to thank the University of Valencia for hospitality.
The research of A.J.M. is supported in part by a grant from PPARC,
U.K..


\end{document}